\documentclass{emulateapj}
\usepackage{apjfonts}
\usepackage[]{natbib}
\usepackage{graphics}
\usepackage{color}

\newcommand{\per}{\ensuremath{^{-1}}}
\newcommand{\persq}{\ensuremath{^{-2}}}
\newcommand{\hal}{H\ensuremath{\alpha}}

\newcommand{\hst}{\emph{HST}}
\newcommand{\msun}{\ensuremath{{M}_{\odot}}}
 
\newcommand{\kms}{km~s\ensuremath{^{-1}}}

\newcommand{\mbh}{\ensuremath{M_\mathrm{BH}}}
\newcommand{\chisq}{\ensuremath{\chi^2}}

\newcommand{\chisqdof}{\ensuremath{\chi^2_{\nu}}}
\newcommand{\sigmastar}{\ensuremath{\sigma_\star}}
\newcommand{\mbul}{\ensuremath{M_\mathrm{bulge}}}
\newcommand{\msigma}{\ensuremath{\mbh-\sigma}}

\newcommand{\lbul}{\ensuremath{L_{\mathrm{bulge}}}}

\newcommand{\sigmaturb}{\ensuremath{\sigma_\mathrm{turb}}}

\newcommand{\rg}{\ensuremath{r_\mathrm{g}}}
\newcommand{\vlos}{\ensuremath{v_\mathrm{LOS}}}
\newcommand{\sigmalos}{\ensuremath{\sigma_\mathrm{LOS}}}
\newcommand{\ml}{\ensuremath{\Upsilon}}
\newcommand{\vsys}{\ensuremath{v_\mathrm{sys}}}
\newcommand{\fscale}{\ensuremath{f_0}}
\newcommand{\aco}{\ensuremath{\alpha_\mathrm{CO}}}
\newcommand{\vcirc}{\ensuremath{v_\mathrm{circ}}}
\newcommand{\vrot}{\ensuremath{v_\mathrm{rot}}}

\begin{document} 

\title{Toward Precision Black Hole Masses with ALMA: \\
  NGC 1332 as a
  Case Study in Molecular Disk Dynamics}

\author{
  Aaron J. Barth\altaffilmark{1},
  Jeremy Darling\altaffilmark{2},
  Andrew J. Baker\altaffilmark{3},  
  Benjamin D. Boizelle\altaffilmark{1}, \\
  David A. Buote\altaffilmark{1}, 
  Luis C. Ho\altaffilmark{4},
  Jonelle L. Walsh\altaffilmark{5}}

\altaffiltext{1}{Department of Physics and Astronomy, 4129 Frederick
  Reines Hall, University of California, Irvine, CA, 92697-4575, USA;
  barth@uci.edu}

\altaffiltext{2}{Center for Astrophysics and Space Astronomy,
  Department of Astrophysical and Planetary Sciences, University of
  Colorado, 389 UCB, Boulder, CO 80309-0389, USA}

\altaffiltext{3}{Department of Physics and Astronomy, Rutgers, the
  State University of New Jersey, Piscataway, NJ 08854-8019, USA}

\altaffiltext{4}{Kavli Institute for Astronomy and Astrophysics,
  Peking University, Beijing 100871, China; Department of Astronomy,
  School of Physics, Peking University, Beijing 100871, China}

\altaffiltext{5}{George P. and Cynthia Woods Mitchell Institute for
  Fundamental Physics and Astronomy, Department of Physics and
  Astronomy, Texas A\&M University, College Station, TX 77843-4242,
  USA}

\begin{abstract}

We present first results from a program of Atacama Large
Millimeter/submillimeter Array (ALMA) CO(2--1) observations of
circumnuclear gas disks in early-type galaxies. The program was
designed with the goal of detecting gas within the gravitational
sphere of influence of the central black holes.  In NGC 1332, the
0\farcs3-resolution ALMA data reveal CO emission from the highly
inclined ($i \approx 83\arcdeg$) circumnuclear disk, spatially
coincident with the dust disk seen in \emph{Hubble Space Telescope}
images.  The disk exhibits a central upturn in maximum line-of-sight
velocity reaching $\pm500$ \kms\ relative to the systemic velocity,
consistent with the expected signature of rapid rotation around a
supermassive black hole. Rotational broadening and beam smearing
produce complex and asymmetric line profiles near the disk center.  We
constructed dynamical models for the rotating disk and fitted the
modeled CO line profiles directly to the ALMA data cube. Degeneracy
between rotation and turbulent velocity dispersion in the inner disk
precludes the derivation of strong constraints on the black hole mass,
but model fits allowing for a plausible range in the magnitude of the
turbulent dispersion imply a central mass in the range
$\sim(4-8)\times10^8$ \msun. We argue that gas-kinematic observations
resolving the black hole's projected radius of influence along the
disk's minor axis will have the capability to yield black hole mass
measurements that are largely insensitive to systematic uncertainties
in turbulence or in the stellar mass profile. For highly inclined
disks, this is a much more stringent requirement than the usual
sphere-of-influence criterion.

\end{abstract}

\keywords{galaxies: nuclei --- galaxies: bulges --- galaxies:
  individual (NGC 1332) --- galaxies: kinematics and dynamics}

\section{Introduction}

Direct measurement of the mass of a supermassive black hole (BH) in
the center of a galaxy generally requires spatially resolved
observations of tracer particles close enough to the BH that their
orbits are dominated, or at least heavily influenced, by the
gravitational potential of the BH. The ``gold standards'' of BH mass
determinations are measurements for the Milky Way's BH based on
resolved stellar orbits \citep{ghez1998, genzel2000, ghez2008} and for
the BH in NGC 4258 based on motions of H$_2$O megamasers resolved by
very long-baseline interferometry \citep{miyoshi1995}. There are two
primary factors that make these the best measurements of BH
masses. First, the observations are able to resolve kinematics at such
small scales that the gravitational potential is overwhelmingly
dominated by the BH itself. Second, the observations are able to
measure the kinematics of individual test particles in orbit about the
BH, rather than the combined and blended kinematics of a population of
tracers having different orbital trajectories.

Aside from the special cases of the Milky Way, NGC 4258, and a small
number of other megamaser disk galaxies \citep[e.g.][]{kuo2011}, most
of the $\sim100$ dynamical detections of BHs come from observations
and modeling of spatially resolved stellar or gas kinematics, mostly
from the \emph{Hubble Space Telescope} (\hst) or large ground-based
telescopes equipped with adaptive optics; for a review of methods and
results see \citet{kormendyho}. Due to the limitations of angular
resolution, these observations do not isolate individual test
particles; rather, they rely on the combined line-of sight kinematics
of stars in the galaxy's bulge, or of gas clouds in a rotating disk,
in both cases modified by the blurring effect of the instrumental
point-spread function.  The stellar-dynamical method is most widely
applicable, in that stars are always present as dynamical tracers in
galaxy nuclei, but the construction of orbit-based dynamical models is
a formidable challenge. Models have recently evolved toward greater
complexity due to a growing recognition that the derived BH masses can
be sensitive to the treatment of the dark matter halo
\citep{gebhardtthomas2009}, triaxial structure
\citep{vandenbosch2010}, and stellar mass-to-light ratio gradients in
the host galaxy \citep{mcconnell2013}.

Measurements of BH mass from ionized gas kinematics are conceptually
and technically simpler since the method relies on modeling circular
rotation of a thin disk rather than modeling the full stellar orbital
structure of an entire galaxy, and early measurements done with
\hst\ spectroscopy demonstrated the potential of this technique
\citep{harms1994, ferrarese1996}.  In contrast to stellar dynamics,
the dynamics of a thin circular disk at small radii are relatively
insensitive to the galaxy's dark matter halo or to stellar
mass-to-light gradients.  However, ionized gas-dynamical measurements
suffer from a separate and significant set of systematic
uncertainties.  Ionized gas disks often have a substantial turbulent
velocity dispersion (\sigmaturb), sometimes up to hundreds of
\kms\ \citep{vandermarel1998, barth2001, verdoeskleijn2006,
  walsh2010}, which must be accounted for in modeling the disk
dynamics. In some cases, the dynamical effect of this turbulence can
affect the estimated BH mass at the factor of $\sim2$ level compared
with masses inferred from thin disk modeling if turbulent pressure
support is neglected. Incorporating turbulent pressure support has
been done using approximations based on point-particle dynamics,
either by applying the formalism of asymmetric drift \citep{barth2001}
or by applying the Jeans equation to model a kinematically hot,
vertically extended rotating disk \citep{neumayer2007}. These methods
are not intrinsically well-suited to gas-dynamical systems, but more
rigorous approaches are still lacking.

Another source of systematic uncertainty is the extended mass
distribution of stars: when the BH's gravitational radius of influence
(\rg; the radius within which $\mbh > 0.5M_\mathrm{total}$) is not
highly resolved, errors in determination of the stellar mass profile
can strongly impact the accuracy of \mbh\ measurements. In many
gas-dynamical measurements done with \hst, \rg\ has been just
marginally resolved, and only in the very best cases such as M87
\citep{macchetto1997, walsh2013} is \rg\ so highly resolved that the
BH dominates the mass profile at the smallest observed scales. This
problem is exacerbated by the very optically thick dust present in
most gas-dynamical targets, which impedes the measurement of the
intrinsic stellar luminosity profile. Despite much early enthusiasm
and a large number of \hst\ orbits invested in the method, ionized gas
dynamics has not produced a very large number of robust BH masses, in
part due to the fact that many galaxies targeted for
\hst\ observations did not exhibit clean rotational kinematics in the
circumnuclear gas \citep{ho2002, hughes2003, noelstorr2007,
  walsh2008}.

In some galaxies, near-infrared ro-vibrational H$_2$ emission lines
from circumnuclear disks of warm molecular gas can be detected.  With
adaptive optics, it is possible to map the kinematics of H$_2$ disks
at high resolution and constrain BH masses
\citep[e.g.,][]{neumayer2007, hicks2008, scharwachter2013,
  denbrok2015}. However, the warm molecular gas in active galactic
nuclei often appears to be somewhat kinematically disturbed or
irregular, even when the gas is in overall rotation about the BH
\citep{mazzalay2014}, and it has not generally been possible to derive
highly precise constraints on BH masses using it as a tracer.

Cold molecular gas has the potential to emerge as an important new
dynamical tracer for BH mass measurement, enabled largely by the
recent construction of the Atacama Large Millimeter/submillimeter
Array (ALMA). The basic principles of BH mass measurement via
molecular gas dynamics are essentially identical to those used for
ionized gas-dynamical measurements, but cold molecular gas offers
several key advantages in terms of the physical structure of
circumnuclear disks and the practical aspects of carrying out a
dynamical measurement.  Similar to ionized gas disks, the dynamics of
rotating molecular disks close to and within \rg\ are essentially
unaffected by the host galaxy's large-scale dark matter halo or
possible triaxial structure, thus avoiding two of the most significant
uncertainties associated with stellar-dynamical BH detection in
early-type galaxies (ETGs).  Crucially, cold molecular gas tends to
have a smaller turbulent velocity dispersion than the ionized gas in
the same galaxy, making it a better tracer of the circular velocity
\citep{young2008, davis2013a} and more amenable to accurate dynamical
modeling.

About 10\% of ETGs contain well-defined, regular, flat, and round
circumnuclear dust disks that can be seen easily in \hst\ images
\citep[e.g.,][]{vandokkum1995, tran2001, laine2003, lauer2005}. In
cases where these disks have associated ionized components, they have
been targets for gas-dynamical BH mass measurements with \hst, but
some dust disks have very weak or undetectable optical line
emission. The optical morphology can provide a clear indication of
dense gas in rotation about the galaxy center, and such disks are
potentially the best and most promising targets for measurement of BH
masses with ALMA.  Pre-ALMA CO observations of molecular gas disks in
ETGs illustrated that well-defined, round dust disks are generally
associated with regular, circular rotation in the molecular gas,
although interferometric observations were not generally able to probe
angular scales as small as \rg\ in nearby galaxies
\citep[e.g.,][]{okuda2005, young2008, alatalo2013}.  The first proof
of concept for BH mass measurement via CO kinematics was presented by
\citet{davis2013b}, who used the Combined Array for Research in
Millimeter-wave Astronomy (CARMA) to observe NGC 4526 at 0\farcs25
resolution, just sufficient to resolve \rg. ALMA now offers the
possibility of routinely carrying out molecular-line observations that
resolve \rg, opening up a major new avenue for determination of BH
masses in galaxy nuclei.

While it has long been anticipated that ALMA will enable BH mass
measurements based on spatially resolved molecular-line kinematics
\cite[e.g.,][]{maiolino2008}, it is not yet clear how widely
applicable this new method will prove to be. The potential of the
method depends crucially on the ability to identify targets having
cleanly rotating gas within \rg, with a high enough surface brightness
for molecular line emission on these scales to be detected and mapped
in reasonable exposure times.  In ALMA Cycle 2, we began a program of
observations of ETGs in order to pursue the goal of obtaining BH mass
measurements. The first stage of this program involves observations of
the CO(2--1) line at $\sim0\farcs3$ resolution to search for evidence
of rapid rotation within the BH sphere of influence. Targets were
selected based on the presence of well-defined, round circumnuclear
dust disks as seen in \hst\ images.  For galaxies showing regularly
rotating disks with high-velocity CO emission from within \rg, deeper
and higher-resolution observations can then be proposed in order to
obtain highly precise measurements of \mbh.

A total of seven ETGs were observed in our Cycle 2 programs, and
observations for the full sample will be presented in a forthcoming
paper.  Here, we present the first results from this project, an
examination of the circumnuclear disk kinematics in NGC 1332.  NGC
1332 is an S0 or E galaxy with a bulge stellar velocity dispersion of
328 \kms\ \citep[for a detailed description of its morphology and
  classification see][]{kormendyho}. It contains a highly inclined and
opaque circumnuclear dust disk that is visible in \hst\ images.  The
BH mass in NGC 1332 has previously been measured using two different
techniques.  \citet{rusli2011} found $\mbh = (1.45\pm0.20)\times10^9$
\msun\ (at 68\% confidence) from stellar-dynamical modeling of VLT
adaptive-optics data.  \citet{humphrey2009} modeled the hydrostatic
equilibrium of the X-ray emitting gas in NGC 1332 using data from the
\emph{Chandra X-ray Observatory} and derived a smaller central mass of
$\mbh = 0.52_{-0.28}^{+0.41} \times10^9$ \msun\ (at 90\%
confidence). The discrepancy between these two measurements provides
additional motivation to attempt to measure \mbh\ via molecular gas
dynamics.  While the nearly edge-on disk inclination makes NGC 1332 a
challenging target for gas-dynamical studies, it is a rare example of
a galaxy that can serve as a test case for comparison of three
independent methods to measure its BH mass, making it a compelling
target for ALMA observations. Following the detection of CO emission
from within \rg\ in these Cycle 2 data, higher-resolution (0\farcs04)
observations of NGC 1332 have been approved for ALMA's Cycle 3.  This
paper provides an initial look at the circumnuclear disk kinematics in
this galaxy at the 0\farcs3 resolution of the Cycle 2 data. This
resolution is sufficient to resolve the BH's radius of influence if
$\mbh > 8\times10^8$ \msun.

The recession velocity of NGC 1332 as measured from optical data is
$1550\pm29$ \kms\ \citep{dacosta1991}. We adopt a distance of 22.3
Mpc for consistency with \citet{rusli2011}, while \citet{humphrey2009}
used a slightly smaller distance of 21.3 Mpc.

\section{Observations}

\begin{figure*}
  \begin{center}
  \includegraphics{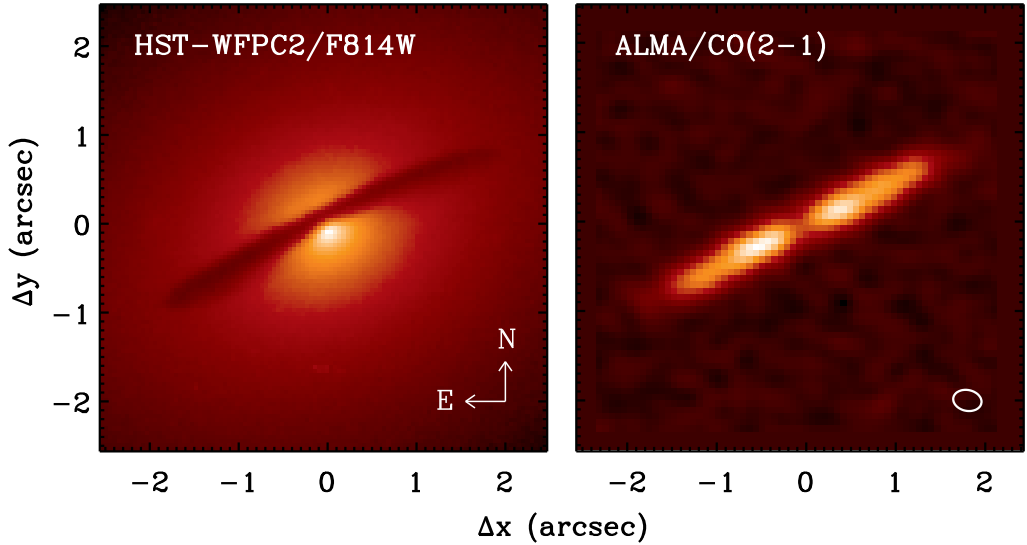}
  \end{center}
  \caption{\emph{Left:} \hst\ WFPC2 F814W ($I$-band) image of the NGC
    1332 nucleus and dust disk, displayed with logarithmic stretch.
    \emph{Right:} ALMA CO(2-1) image, summed over all frequency
    channels of the data cube and displayed with linear stretch. The
    ellipse at lower right illustrates the FWHM size of the ALMA
    synthesized beam.
  \label{hstalmafig}}
\end{figure*}

As part of Program 2013.1.0229.S, ALMA observed NGC 1332 on 2014
September 1 in a frequency band centered on the redshifted
$^{12}$CO(2-1) 230.538 GHz line at $\nu_\mathrm{obs} = 229.37$ GHz (in
ALMA Band 6). The minimum and maximum baselines of the array were 33.7
m and 1.1 km, respectively, and the on-source integration time was
22.7 min. Observations were processed using version 4.2.2, r30721 of
the Common Astronomy Software Application
\citep[CASA;][]{mcmullin2007} package and version 31266 of the
standard ALMA pipeline.  This produced a data cube spanning a field of
view of 21\arcsec$\times$21\arcsec\ with 0.07\arcsec\ spatial pixels
(corresponding to 7.6 pc pixel\per), and a separate continuum image
having the same pixel scale.  Use of Briggs weighting with robustness
parameter 0.5 yields a synthesized beam with major and minor axis FWHM
sizes of $0\farcs319$ and $0\farcs233$ and major axis position angle
78\fdg4, giving a geometric mean resolution of 0\farcs27 corresponding
to 29.2 pc in NGC 1332.  The data cube contains 60 frequency channels
with spacing 15.4 MHz, or velocity spacing 20.1 \kms\ relative to the
frequency of the CO(2--1) line at the NGC 1332 systemic velocity.  The
RMS noise level as measured in line-free regions of the data cube is
0.4 mJy beam\per\ per channel.  The overall flux scale for the data
was set by an observation of the quasar J0334$-$4008, for which we
adopt a 10\% uncertainty \citep[see, e.g., the Appendix
  of][]{alma2015} in the pipeline-reported value of 0.76\,Jy at
234.2\,GHz.

Line emission is visible in 52 channels in the data cube, spanning a
full velocity width of 1040 \kms.  Figure \ref{hstalmafig} shows the
ALMA CO image of NGC 1332, summed over all frequency channels of the
data cube.  CO emission originates from a region corresponding closely
to the optically thick dust disk with 2\farcs2 angular radius, as seen
in an archival \hst\ WFPC2 F814W ($I$-band) image. There is no
significant CO emission detected above the level of the background
noise at any location outside the circumnuclear disk. The disk's CO
surface brightness distribution exhibits a dip within the innermost
$r\lesssim0\farcs2$, but even in this region CO emission is clearly
detected.  The continuum image reveals a marginally resolved source at
the disk center with flux density $8.75 \pm 0.04$ mJy; we assume an
additional 10\% uncertainty in this value based on the uncertainty in
the overall flux calibration.

\section{CO emission properties and disk kinematics}

\begin{figure}
  \scalebox{0.42}{\includegraphics{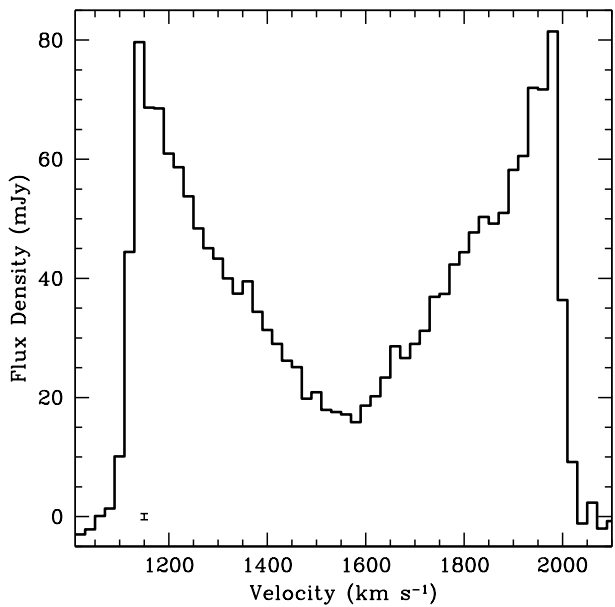}}
  \caption{CO(2--1) line profile integrated over an elliptical region
    having major and minor axis lengths of $4\farcs3$ and
    $0\farcs7$. The error bar at lower left shows the background noise
    level summed in quadrature over this integration region.
    \label{lineprofilefig}}
\end{figure}

Figure \ref{lineprofilefig} illustrates the integrated CO emission
profile over this region, which exhibits a symmetric double-horned
shape with peaks separated by 840 \kms. To examine the spatially
resolved kinematics, we fit the line profile at each spatial pixel
using a sixth-order Gauss-Hermite function \citep{vandermarel1994}.
The S/N was sufficient to obtain successful fits at each pixel over an
elliptical region having major and minor axis lengths of 4\farcs3 and
0\farcs7.  Outside this region, the CO flux drops rapidly to below the
level of the noise.

\begin{figure*}
  \begin{center}
    \scalebox{1.0}{\includegraphics{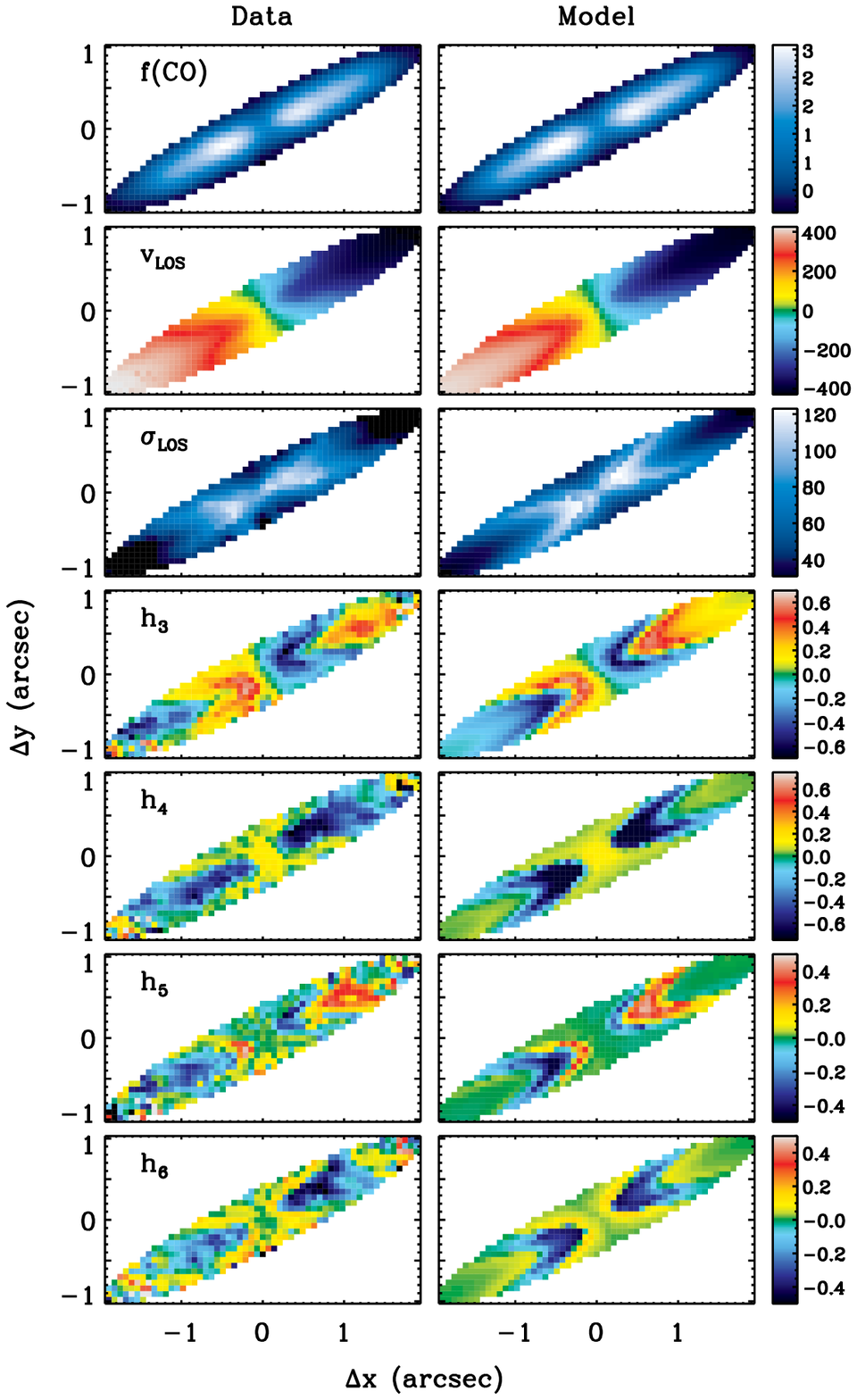}}
    \vspace*{0.2in}
  \end{center}
  \caption{Maps of CO intensity, \vlos, \sigmalos, $h_3$, $h_4$,
    $h_5$, and $h_6$ measured from Gauss-Hermite fits to the data and
    model cubes. The left panels show kinematic moment maps of the
    ALMA data, and the right panels show the best-fitting model for
    the flat \sigmaturb\ profile having $\mbh=6.0\times10^8$ \msun, as
    described in \S\ref{sec:flatsigma}. The model CO intensity map
    gives the integral of the CO line profile at each spatial pixel in
    the modeled data cube. Units for the CO surface brightness map are
    Jy beam\per\ \kms, and units for the \vlos\ and \sigmalos\ maps
    are \kms. In this and subsequent figures, line-of-sight velocities
    are shown relative to the systemic velocity 1559 \kms\ determined
    from the best-fitting model. The higher-order moment coefficients
    $h_3$ through $h_6$ are dimensionless.
    \label{momentmaps}}
\end{figure*}

Figure \ref{momentmaps} shows the spatially resolved kinematic moment
maps for the line-of-sight velocity centroid (\vlos, measured relative
to the systemic velocity), the dispersion \sigmalos, and the
higher-order Gauss-Hermite moments $h_3$ through $h_6$. The
odd-numbered moments $h_3$ and $h_5$ describe asymmetric departures
from a Gaussian profile, while the even-numbered moments $h_4$ and
$h_6$ quantify symmetric deviations from a Gaussian shape.  The
\vlos\ map illustrates the fact that the NGC 1332 disk is in orderly
rotation overall.  Beam smearing of the disk's velocity field is
particularly severe given the disk's nearly edge-on inclination.  At
0\farcs3 resolution, the disk's projected semi-minor axis is
essentially unresolved, and low-velocity emission ``piles up'' into
the line profiles along the disk major axis. Consequently,
high-velocity emission from gas in rotation near the disk center is
spatially blended with a substantial amount of low-velocity emission
from foreground and background regions along the disk surface.

The \vlos\ map for NGC 1332 shows a relatively steep velocity gradient
across the nucleus from the southeastern (redshifted) to northwestern
(blueshifted) side of the disk, but without central high-velocity
emission. High-velocity emission is present near the disk center, but
as a sub-dominant contribution to the line profiles.  The primary
signature of the disk's central rise in rotation speed is in the $h_3$
map, which shows very strong blueward and redward asymmetries on
opposite sides of the disk center. The \sigmalos\ map has an ``X''
shape typical of inclined rotating disks, with peak line-of-sight
velocity dispersion $\sim120$ \kms. This ``X'' shape comes from
rotational broadening and beam smearing in regions of the disk
exhibiting a steep gradient in line-of-sight velocity. It is also
possible that a portion of the spatial variation in observed line
width could be due to a radial gradient in the disk's intrinsic
turbulent velocity dispersion, but dynamical modeling is required to
distinguish the separate contributions of beam smearing and turbulent
velocity gradients. The $h_5$ and $h_6$ maps are noisy but still
contain coherent structure tracing the same features as $h_3$ and
$h_4$.

To examine the kinematic position angle (PA) as a function of radius
in the disk, we used the kinemetry method of \citet{krajnovic2006},
applying kinemetry to the measured \vlos\ map. The kinemetry routine
fits a harmonic expansion to \vlos\ along elliptical annuli. Its
output includes, at each semi-major axis distance $R$, the kinematic
position angle $\Gamma$, the axis ratio $q$ of the kinematic ellipse,
the first harmonic coefficient $k_1$ (equivalent to the major axis
line-of-sight velocity amplitude), and the ratio $k_5/k_1$. The $k_5$
coefficient quantifies deviations from pure rotation, and the presence
of strong features in the $k_5/k_1$ map can indicate the existence of
multiple kinematic components. The kinematic PA $\Gamma$ refers to the
kinematic major axis and is defined here such that PA=0\arcdeg\ would
correspond to a north-south orientation for the disk major axis with
the northern side of the disk redshifted. We use the term kinematic
major axis to refer to the locus of points of maximum line-of-sight
velocity amplitude on each elliptical annulus, which is equivalent to
the line of nodes for an inclined, flat circular disk seen in
projection. The kinematic minor axis refers to the locus of points at
which \vlos\ is equal to the systemic velocity \vsys. In a flat
circular disk, the kinematic major and minor axes will be seen as
orthogonal straight lines on the sky, but in the presence of warping,
radial flows, or elliptical orbits, this is in general no longer the
case. (See Wong, Blitz, \& Bosma 2004 for a detailed discussion of the
observable signatures of noncircular or warped kinematics in gaseous
disks.)

As described by \citet{krajnovic2008}, galaxies having multiple
kinematic components are typically characterized by either abrupt
changes in $q$ or $\Gamma$ (with $\Delta q > 0.1$ or $\Delta\Gamma >
10\arcdeg$), or a double-peaked $k_1$ profile, or the presence of a
distinct peak in the $k_5$ profile at which $k_5/k_1$ > 0.02. A
kinematic twist is identified if $\Gamma$ varies smoothly with radius
with a rotation of $>10\arcdeg$.

\begin{figure}
  \scalebox{0.4}{\includegraphics{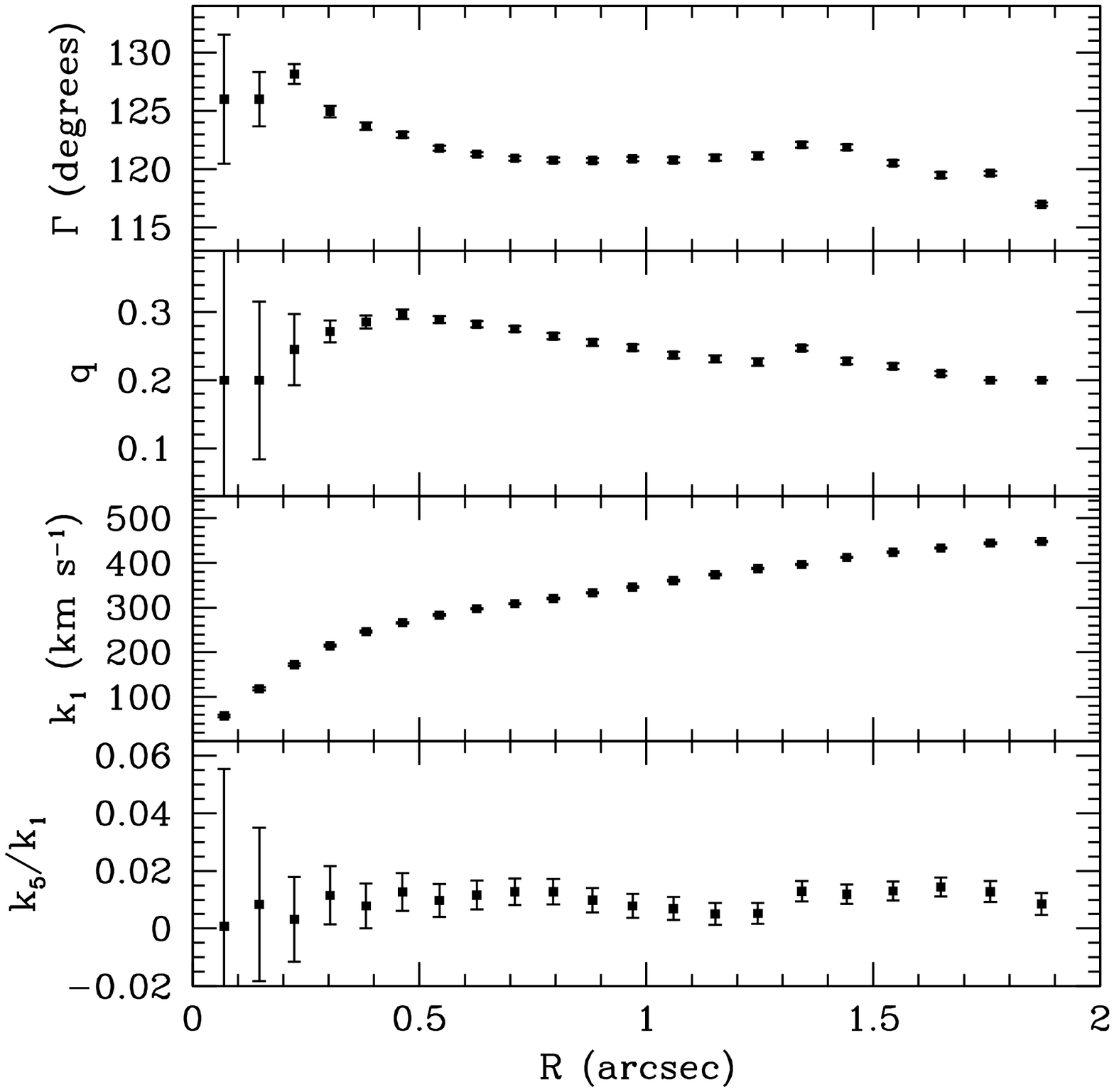}}
  \caption{Kinemetry for the NGC 1332 line-of-sight velocity (\vlos)
    map. From top to bottom, the panels illustrate the kinematic PA
    $\Gamma$, the flattening $q$, the $k_1$ coefficient (equivalent to
    the amplitude of \vlos\ along the disk major axis), and the ratio
    $k_5/k_1$.}
  \label{fig:kinemetry}
\end{figure}

The kinemetry results are shown in Figure \ref{fig:kinemetry}.  The
kinematic PA $\Gamma$ ranges from 128\arcdeg\ at $r=0\farcs22$ (within
one resolution element of the galaxy nucleus) to 117\arcdeg\ at the
outer edge of the disk, which can be characterized as a mild kinematic
twist.  The $k_5/k_1$ ratio is below 0.02 at all radii and does not
contain any strong features. Overall, the modest radial variation in
$\Gamma$ and $q$, and the flat $k_5/k_1$ profile with magnitude of
$\sim1\%$, indicate that the NGC 1332 disk kinematics are consistent
with coherent disk rotation as the dominant kinematic structure.  If
\rg\ were well resolved, $k_1$ would exhibit a central
quasi-Keplerian ($v\sim r^{-0.5}$) decline with increasing
distance from the nucleus, but instead we see a smooth monotonic rise
in $k_1$ as a function of $R$. This is an indication that \vlos\ does
not in itself provide direct evidence for a compact central mass; beam
smearing hides the central high-velocity emission in the higher-order
velocity moment maps.

\begin{figure}
\scalebox{0.5}{\includegraphics{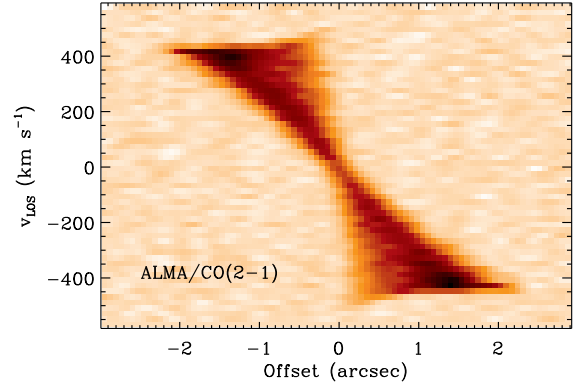}}
\caption{Position-velocity diagram along the disk major axis, from a
  four-pixel extraction width. Data are displayed with a linear
  stretch. }
\label{pvdfig}
\end{figure}

We also extract a position-velocity diagram (PVD) as another method to
visualize the kinematics. We constructed the PVD by rotating each
frequency slice of the data cube clockwise by 27\arcdeg\ so that the
kinematic PA at the largest measured radius was oriented horizontally,
and then extracting a four-pixel wide swath through the data cube
(corresponding to one resolution element) along the disk major
axis. Figure \ref{pvdfig} shows the PVD, which illustrates a smooth
and continuous distribution of emission across the full range of
velocities present in the disk. The central high-velocity emission is
clearly seen in the form of a slight upturn in the locus of maximum
line-of-sight speed on either side of the nucleus. This central
velocity upturn is relatively faint and only seen within $\pm0\farcs8$
of the nucleus, it only spans 3--4 velocity channels above the level
of the flat outer envelope in rotation velocity, and it is spatially
blended with a much larger amount of emission spanning the entire
velocity range present in the disk.  This PVD structure may be
contrasted with the case of an observation in which \rg\ is well
resolved: in that situation, the central velocity upturn would be seen
as a narrow locus of emission, rather than as the outer envelope of a
broad distribution of velocities extending down to zero. Nevertheless,
the velocity rise seen within the innermost $\sim0\farcs5$ is the
expected signature of a compact central massive object that dominates
the gravitational potential at small radii. Although the CO surface
brightness map shows a central dip in brightness within the innermost
$r<0\farcs2$, the observed velocity structure provides evidence that
the intrinsic CO-emitting region extends to scales within the black
hole's sphere of influence.

From the PVD, we can extract an estimate of the enclosed gravitating
mass within one resolution element of the galaxy nucleus. At a
distance of four pixels (0\farcs28) from the galaxy center,
corresponding to $r=30$ pc, the maximum velocity seen in the PVD is
$\approx480$ \kms. If this maximum velocity is equated with the
circular speed at $r=30$ pc, the implied enclosed mass is $M(r)
\approx 1.6\times10^9$ \msun, which can then be taken as an upper
limit to \mbh\ since it includes both the BH and the stellar mass
enclosed within this radius. A caveat to this simple estimate is that
the CO line profiles may be broadened by turbulence in the molecular
gas disk, in which case the outermost velocity envelope seen in the
PVD would give an upper limit to the line-of-sight circular velocity.

The PVD structure also helps to clarify why the integrated CO line
profile in Figure \ref{lineprofilefig} is so strongly double
peaked. The flat outer velocity envelope to the PVD at
$r\sim0\farcs8-2\farcs2$ indicates that the galaxy's rotation curve is
fairly flat over this range of radii. As a result, there is a large
amount of CO emission at line-of-sight velocities in a narrow range
around $\pm400$ \kms\ from gas along the disk's major axis.

The total CO(2-1) flux of the disk is $37.7\pm0.3$ Jy \kms, with an
additional 10\% uncertainty in the flux scale. We can use this flux
value to obtain an estimate of the total molecular gas mass of the
disk by applying a CO-to-H$_2$ conversion factor $\aco$
\citep[e.g.,][]{bolatto2013}.  Following \citet{carilli2013}, we
convert from observed flux to luminosity to obtain
$L^{\prime}_\mathrm{CO(2-1)} = (1.14\pm0.11)\times10^7$ K
\kms\ pc$^{2}$.  The CO-to-H$_2$ conversion is most reliably
calibrated in terms of the luminosity of the CO(1--0) emission line,
which we do not have for NGC 1332. Following \citet{sandstrom2013}, we
assume an intensity ratio $R_{21} \equiv
L^{\prime}_\mathrm{CO(2-1)}/L^{\prime}_\mathrm{CO(1-0)} = 0.7$.  For
the sample of 18 ETGs studied by \citet{crocker2012}, the median CO
intensity ratio is $R_{21} = 0.76$ with standard deviation of 0.23,
consistent with our adopted value of 0.7.  Then, a standard value of
$\aco = 3.1$ \msun\ pc\persq\ (K \kms)\per\ derived for nearby
galaxies and including a factor of 1.36 correction for helium
\citep{sandstrom2013} leads to an estimated molecular gas mass of
$M_\mathrm{gas} = \aco L^{\prime}_\mathrm{CO(1-0)} =
(5.0\pm0.5)\times10^7$ \msun.  \citet{sandstrom2013} find that within
the central kpc of galaxies, \aco\ is typically a factor of $\sim2$
below this adopted standard value, however, and in some cases the
value of \aco\ in galaxy centers is an order of magnitude lower than
the average. If this trend applies to ETGs as well, then the actual
gas mass in the disk could be several times lower than this simple
estimate. The circumnuclear disk of NGC 1332 may have physical
conditions very different from the environments in which \aco\ has
been calibrated, and in regions of high density and high optical depth
the standard assumptions of the CO-to-H$_2$ conversion method may be
invalid \citep{bolatto2013}. We therefore consider the gas mass
derived above as merely a rough estimate.

\section{Dynamical Modeling}

\subsection{Method}

We modeled the kinematics of a flat circular disk following the same
approach used for ionized gas disks observed with
\hst\ \citep{macchetto1997, barth2001, walsh2010}. The calculation
begins with a mass model consisting of the central point mass
\mbh\ and the extended mass profile of stars in the host galaxy
$M_\star(r)$, representing the mass enclosed within radius $r$. The
mass profile $M_\star(r)$ is determined from an empirically measured
and deprojected luminosity profile, with the stellar mass-to-light
ratio \ml\ as a free parameter. 

We neglect the possible contribution of dark matter, which is expected
to be extremely small relative to stellar mass on scales comparable to
the circumnuclear disk radius. The mass model of NGC 1332 from
\citet{humphrey2009} indicates a total enclosed dark matter mass of
$\approx10^8$ \msun\ within $r<250$ pc, two orders of magnitude lower
than the stellar mass enclosed at this scale.  Our model does not
include the mass of the gas disk itself, since our estimate of the
total molecular gas mass is $\sim5\times10^7$ \msun\ within $r<250$
pc.

A circular, flat disk in rotation about the galaxy center has rotation
speed given by $v_\mathrm{rot}(r) = [GM(r)/r]^{1/2}$ where $M(r) =
\mbh + M_\star(r)$.  The line-of-sight velocity at each point in the
model disk is then determined for a given inclination angle $i$ and a
major axis orientation angle $\Gamma$ \citep[for details
  see][]{macchetto1997, barth2001}. The model is constructed on a
spatial grid that is highly oversampled relative to the ALMA spatial
pixel size, because a given pixel near the disk center can contain gas
spanning a substantial velocity range. In the models, each ALMA pixel
is subdivided into an $s\times s$ grid of sub-pixel elements. At each
sub-pixel element in the grid, we model the emergent line profile as a
Gaussian, with central velocity given by the projected rotation
velocity at that point, and with some turbulent linewidth \sigmaturb.
We discuss the possible radial variation of \sigmaturb\ in
\S\ref{sec:turbmodel}.

The line profiles at each location must be weighted by the CO surface
brightness at the corresponding point in the disk. The available
information on the CO surface brightness distribution of the disk is
simply the velocity-integrated CO image as shown in Figure
\ref{hstalmafig}.  However, this is an image of the disk modified by
beam smearing, and the model requires an image of the intrinsic
surface brightness distribution. Ideally one would want a CO surface
brightness map at the same resolution as the $s\times s$ sub-pixel
sampling of the model grid, but such information is not available.
Lacking knowledge of the CO surface brightness distribution on
subpixel scales, the line profiles at each sub-pixel grid point
contributing to a given ALMA spatial pixel were normalized to have
equal fluxes. Then, the subsampled data cube was rebinned to the
spatial pixel scale of the ALMA data by averaging each $s\times s$ set
of sub-pixel line profiles to form a single profile.  The line
profiles are calculated on a grid of 20.1 km s\per\ pixel\per,
corresponding to the velocity sampling of the ALMA data cube.

In order to scale the line profile of each spatial pixel to its
appropriate flux level, we used a deconvolved CO surface brightness
map. We deconvolved the CO image shown in Figure \ref{hstalmafig}
using five iterations of the Richardson-Lucy algorithm
\citep{richardson1972, lucy1974} implemented in IRAF\footnote{IRAF is
  distributed by the National Optical Astronomy Observatories, which
  are operated by the Association of Universities for Research in
  Astronomy, Inc., under cooperative agreement with the National
  Science Foundation.}, where the deconvolution was carried out with
an elliptical Gaussian point-spread function (PSF) matching the
specifications of the ALMA synthesized beam. The choice of five
iterations of the deconvolution algorithm is somewhat arbitrary: this
produced an adequately sharpened image, and larger numbers of
iterations began to produce noticeable artifacts by amplifying
background noise. In the model calculation, the line profile at each
pixel was then normalized to match the flux at the corresponding pixel
in the deconvolved flux map. To allow for any possible mismatch in
flux normalization between the deconvolved flux map and the original
ALMA data cube, we multiply the flux map by a scaling factor
\fscale\ that is a free parameter in the model fits. In practice, the
best-fitting value of \fscale\ is very close to unity.

Each frequency slice of the modeled cube is then convolved with the
ALMA synthesized beam, modeled as an elliptical Gaussian, producing a
simulated data cube analogous to the ALMA data.  Since we rebin the
high-resolution model cube to the ALMA pixel resolution prior to
carrying out the PSF convolution, there is little time penalty for
carrying out the initial computation of line profiles on an
oversampled pixel grid with $s$ as high as 10, and it is feasible to
carry out model optimizations with the line profiles computed at much
finer spatial sampling ($s=50$ for example). PSF convolution is often
the most time-consuming step of gas-dynamical model calculations,
particularly if the convolution is carried out on an oversampled pixel
grid.

Free parameters in the model include \mbh\ and \ml, the $x$ and $y$
centroid positions of the BH, the systemic velocity \vsys, the
inclination and orientation angles of the disk $i$ and $\Gamma$, the
flux-normalization factor \fscale\ applied to the CO flux map, and the
parameters describing the run of \sigmaturb\ as a function of radius
in the disk (see \S\ref{sec:turbmodel}; this requires up to three
parameters depending on the choice of \sigmaturb\ model).

\subsection{Stellar Mass Profile}

The stellar mass profile $M_\star(r)$ is an essential ingredient in
the dynamical modeling, and is determined by measuring and
deprojecting the galaxy's surface brightness profile. Ideally, this
should be measured from imaging data having angular resolution at
least as high as the spectroscopic data. Near-infrared observations
are strongly preferred, especially for galaxies having dusty
nuclei. The only available \hst\ image of NGC 1332 is the WFPC2 F814W
image shown in Figure \ref{hstalmafig}, and the dust disk is extremely
optically thick in the $I$ band.

\citet{rusli2011} used a combination of seeing-limited $R$-band data
on large scales, and $K$-band AO data on small scales ($<4\farcs5$),
to measure the stellar luminosity profile of NGC 1332; they masked out
the central dust disk in extracting the galaxy's light profile. Their
model was then deprojected under the assumption of axisymmetric
structure and an inclination of 90\arcdeg. We used this same mass
profile (kindly provided by J.\ Thomas) in order to carry out a direct
comparison between our gas-dynamical modeling and the
stellar-dynamical results of \citet{rusli2011}. The version of the
mass profile provided (which we denote $M_{R\star}$) was the average
of the best-fitting stellar-dynamical models over the four quadrants
of the VLT integral-field kinematic data, and corresponds to a stellar
mass-to-light ratio in the $R$ band of $\Upsilon_R = 7.35$. In our
models, we incorporated this mass profile directly, multiplied by a
mass-to-light scaling factor $\gamma$ (a free parameter in the model
fits), such that $M_\star(r) = \gamma M_{R\star}(r)$ and the resulting
mass-to-light ratio is $\Upsilon_R = 7.35\gamma$. As described below,
our best fits converged on values of $\gamma$ very close to unity,
indicating that our gas-dynamical model fitting finds close
agreement with the stellar mass distribution measured by
\citet{rusli2011}.

As a consistency check, we also measured a stellar luminosity profile
from the archival \hst\ data. The \hst\ observation was taken as two
individual 320 s exposures with the galaxy nucleus on the PC chip of
the WFPC2 camera (0\farcs0456 arcsec pixel\per), and we combined the
two exposures using the IRAF/STSDAS task \texttt{crrej}. We measured
the galaxy's surface brightness profile from the PC image, first
masking point sources, globular clusters, and the dust disk. The disk
was masked over the radial range 0\farcs1--2\farcs3. The central four
pixels, where dust extinction appears less severe, were retained in
order to anchor the radial profile measurement at the galaxy's
nucleus, but the measurement then gives a lower limit to the central
surface brightness due to dust extinction along the nuclear line of
sight.

The surface brightness profile was fit using a 2D multi-Gaussian
expansion (MGE) using the \texttt{MGE\_FIT\_SECTORS} package
\citep{cappellari2002}, using eight Gaussian components. The MGE
expansion requires a PSF model, and we modeled the PC F814W PSF using
an MGE fit to a synthetic PSF generated using Tiny Tim
\citep{krist2011}. Deprojection of the MGE model was done assuming
that the galaxy bulge is an oblate spheroid with inclination
83\arcdeg\ (based on fits to the ALMA kinematics described below). CCD
counts were converted to $I$-band solar units accounting for a
Galactic foreground extinction of $A_I = 0.049$ mag
\citep{schlafly2011} to give the $I$-band luminosity profile $L_I(r)$.
The mass profile is then given by $M_\star(r) = \Upsilon_I L_I(r)$,
where $\Upsilon_I$ is the $I$-band mass-to-light ratio. 

\begin{figure}
\scalebox{0.4}{\includegraphics{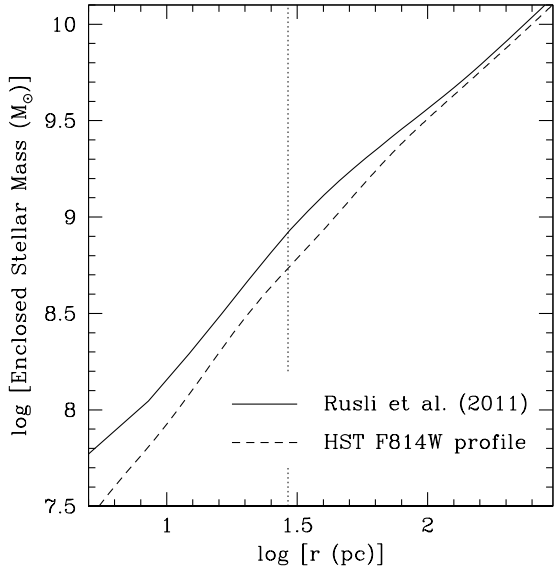}}
\caption{Enclosed stellar mass profiles $M_\star(r)$ as measured by
  \citet{rusli2011} (solid curve) and as measured from the \hst\ WFPC2
  F814W image (dashed curve). The mass-to-light ratio in each case is
  based on the best-fitting dynamical model with a spatially uniform
  value of the turbulent velocity dispersion \sigmaturb. The vertical
  dotted line corresponds to the 0\farcs27 resolution of the ALMA
  data. }
\label{massprofiles}
\end{figure}

Figure \ref{massprofiles} shows the enclosed stellar mass profile
$M_\star(r)$ from \citet{rusli2011} and the profile measured from the
\hst\ F814W image. In both cases, the mass-to-light ratio
normalization is based on the best-fitting result for models assuming
a spatially uniform value of \sigmaturb, as described in
\S\ref{sec:turbmodel}. The $I$-band stellar mass profile implies a
lower stellar mass than the $R$+$K$ band profile from
\citet{rusli2011}. The most plausible explanation is that this mass
deficit is the result of dust absorption by the circumnuclear disk in
the $I$-band \hst\ image, which could not be entirely masked out. We
therefore choose to use the \citet{rusli2011} profile for our final
model fits. At the outer edge of the disk at $r=240$ pc, the enclosed
stellar mass reaches $10^{10}$ \msun, so the BH is expected to
make a small contribution to the total enclosed mass at this
scale.

\subsection{Turbulent Velocity Dispersion Profile}
\label{sec:turbmodel}

To model the CO line profiles, some assumption must be made regarding
the intrinsic velocity dispersion of the gas. The thermal contribution
to the linewidth for cold molecular gas will be extremely small:
\citet{bayet2013} find kinetic temperatures of typically $\sim10-20$ K
for molecular gas in ETG disks.  Ideally, with angular resolution high
enough to resolve individual giant molecular clouds (GMCs) within the
disk, the turbulent velocity dispersion of individual GMCs can be
measured. In a range of environments, the highest linewidths observed
in individual GMCs correspond to $\sigmaturb \approx 30-40$
\kms\ \citep{leroy2015}. The closest analog to NGC 1332 having
observations of high enough resolution to isolate individual GMCs is
NGC 4526, where CO(2--1) observations from CARMA were able to resolve
103 individual GMCs \citep{utomo2015}.  In NGC 4526, most of the
individual GMCs were found to have $\sigmaturb\approx5-10$ \kms, and
the highest-dispersion clouds have $\sigmaturb\approx25$ \kms. For
NGC 1332, the 0\farcs3 resolution of our ALMA data shows a very smooth
distribution of CO emission across the disk surface, and much higher
angular resolution would be required in order to identify individual
clouds within the disk or measure their velocity dispersions directly.

In the outer regions of the NGC 1332 disk, the CO emission profiles
are very narrow. At the largest radii along the disk major axis, the
lines are unresolved as observed in the 20 \kms\ velocity channels,
with measured linewidths of $\sigma \approx 10-20$ \kms\ from the
Gauss-Hermite fits. Toward the inner regions of the disk, the
linewidths rise to a maximum of $\sigma=120$ \kms, with the largest
widths observed at locations about 0\farcs3--0\farcs4 on either side
of the nucleus along the disk major axis. In this central region, the
line profiles are extremely asymmetric with broad, extended wings. As
will be shown below, most or all of this rise in observed linewidth
can be explained by beam smearing of unresolved rotation. There
remains the possibility, however, of a genuine increase in the gas
turbulent velocity dispersion toward the disk center. One issue in the
model construction is that if the pixel oversampling factor $s$ is too
low, the models can have a tendency to require a spuriously
high value of $\sigmaturb$ to compensate for inadequate spatial
sampling of rotational velocity gradients. It is crucial, therefore,
to ensure that the models are computed on a grid with sufficient
spatial sampling that \sigmaturb\ is not biased toward higher values
than actually occur in the disk.

If the molecular gas in the NGC 1332 disk is arranged in unresolved
discrete clouds then the CO line width from an individual cloud is the
result of the internal turbulent velocity dispersion within the cloud,
rotation of the cloud as a whole, and shear due to galactic
differential rotation. Additionally, random motions of clouds (either
in-plane or vertical) will contribute to the observed line widths.
Since our observations do not resolve individual clouds, our data
cannot distinguish among these contributions, and we use the term
``turbulence'' to refer to the combination of all processes
responsible for the emergent line width from a given location at the
disk surface. In their study of the NGC 4526 GMC population,
\citet{utomo2015} found that the energy in turbulent motion dominated
over internal rotational energy for nearly all of the resolved clouds.

To explore how the model fits depend on the assumed turbulent velocity
dispersion profile and allow for possible radial gradients in
turbulence, we ran models using the following prescriptions for
$\sigmaturb(r)$.

\emph{Flat:} The simplest parameterization is \sigmaturb\ = constant,
allowing \sigmaturb\ to be a free parameter in the model fits.

\emph{Exponential:} Following a typical prescription used in ionized
gas dynamics, we used a model of the form $\sigmaturb = \sigma_0
\exp(-r/r_0) + \sigma_1$, where $\sigma_0$, $\sigma_1$, and $r_0$ are
free parameters. Model fits consistently drove the value of $\sigma_1$ to
zero, so this parameter was discarded from final fits. To prevent the
line profile widths from becoming arbitrarily narrow, we enforced a
minimum value of $\sigmaturb=1$ \kms.

\emph{Gaussian:} The largest observed linewidths are seen at locations
slightly offset from the disk center, suggesting that a
\sigmaturb\ model with a central depression could provide a better fit
to the data. As a simple model allowing for a central plateau or dip
in \sigmaturb, we used a Gaussian profile: $\sigmaturb = \sigma_0
\exp[-(r - r_0)^2 / (2\mu^2)] + \sigma_1$, where $\sigma_0$,
$\sigma_1$, $r_0$, and $\mu$ are free parameters. Again, we found that
model fits drove $\sigma_1$ to zero and we removed this parameter from
the final model runs, and enforced a minimum value of $\sigmaturb = 1$
\kms. The parameter $r_0$ was allowed to vary over positive and
negative values for maximum freedom in fitting the data. Positive
values were preferred by the fits, producing a central dip in the
turbulent velocity dispersion.

In \S\ref{sec:results}, we present a comparison of model fitting
results using each of these parameterizations for \sigmaturb.  We
emphasize that none of these prescriptions for \sigmaturb\ represents
a physically motivated model.  The disk's actual \sigmaturb\ profile
could be considerably different from any of these prescriptions (and
might not be axisymmetric), but the data do not provide sufficient
information to justify modeling \sigmaturb\ with anything much more
complex than these simple parameterizations.

\subsection{Model Optimization}

The output of each model computation for a given parameter set is a
simulated data cube having the same spatial and velocity sampling as
the ALMA data. Thus, the models can be fitted directly to the ALMA
data cube to optimize the fit by \chisq\ minimization.  We fit models
using the \texttt{amoeba} downhill simplex method implemented in IDL.

For model fits to the observed data cube, we fit models only to the
spatial pixels within the elliptical region illustrated in Figure
\ref{momentmaps}, using 52 frequency channels at each spatial pixel
spanning the full width of the CO emission line.  To determine
\chisq\ by direct comparison of the modeled cube with the data, an
estimate is required for the flux uncertainty at each pixel in the
data. The simplest way to estimate the noise level would be to
determine the standard deviation of pixel values in emission-free
regions of the data cube. However, the background noise in the data
exhibits strong correlations among neighboring pixels, on angular
scales similar to the synthesized beam size. A proper calculation of
\chisq\ in the presence of strongly correlated errors requires the
computation and inversion of the covariance matrix of the data
uncertainties \citep[e.g.,][]{gould2003}. We attempted to construct a
covariance matrix by using ``blank'', emission-free regions of the
data cube to determine the pixel-to-pixel correlations in the
background noise. Within the elliptical fitting region, there are 523
individual spatial pixels. We found that numerical errors in
constructing and inverting the covariance matrix rendered the
\chisq\ calculation highly unstable.

We then opted for a simplified approach to computing \chisq. The
background noise correlations are strong on scales comparable to the
synthesized beam, of size $\sim0\farcs3$, while the pixel size is
0\farcs07.  We rebinned the data by spatially averaging the flux over
$4\times4$ pixel blocks within each frequency channel, yielding
approximately one rebinned pixel per synthesized beam. This process
averages over the mottled pattern in the background noise seen at the
original pixel resolution, producing a data cube in which the
background noise is nearly uncorrelated among neighboring pixels. In
this block-averaged data cube, we measured the standard deviation of
pixel values in blank regions to determine the noise level for use in
measuring \chisq. The RMS noise level in the $4\times4$ block-averaged
data is 0.3 mJy beam\per\ (compared with 0.4 mJy beam\per\ in the
original data). The fact that the noise in the block-averaged data is
nearly as large as in the full-resolution data (rather than scaling as
$1/\sqrt{n}$ where $n$ is the number of binned pixels) reflects the
strong local correlations in the pixel values of the full-resolution
data.  Then, for each model iteration, the calculated model at the
original ALMA pixel scale was similarly block-averaged over $4\times4$
pixel regions to compare with the block-averaged data. In the rebinned
data, we calculate \chisq\ over 49 block-averaged spatial pixels and
52 frequency channels at each pixel, for a total of 2548 data points.

\section{Modeling Results}
\label{sec:results}

\subsection{Initial tests}

\begin{figure}
  \scalebox{0.4}{\includegraphics{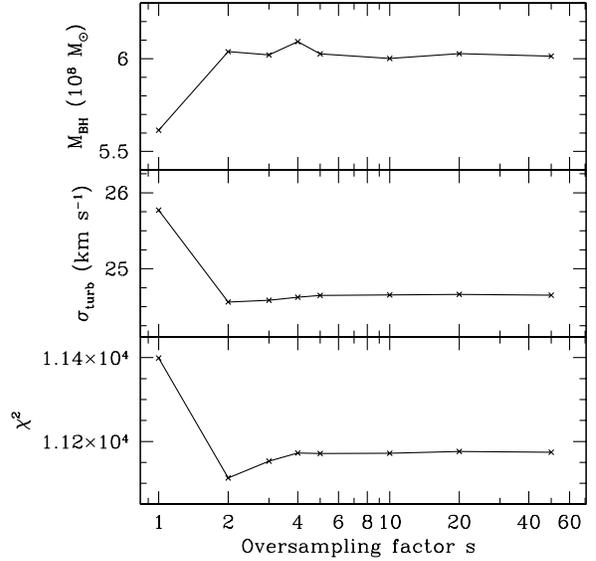}}
  \caption{Best-fitting values of \mbh\ and \sigmaturb, and the
    corresponding \chisq\ values, for models with flat \sigmaturb\ and
    oversampling factor $s$ ranging from 1 to 50. The number of
    degrees of freedom in the fit is 2540. }
  \label{subsampfig}
\end{figure}

As an initial test to examine the impact of different values of the
oversampling factor $s$, we ran fits with models having a flat
turbulent velocity dispersion and oversampling factors of $s=1$, 2, 3,
4, 5, 10, 20, and 50. All model parameters were free in these initial
fits. Figure \ref{subsampfig} illustrates the results of varying
$s$. An oversampling factor of $s=2$ was the bare minimum required in
order to obtain modeled P--V diagrams having smooth structure similar
to the observed PVD.  The models converged on consistent results for
$\chisq$ and for the free parameters when the oversampling factor was
greater than $s=4$. We choose $s=10$ for our final model fits since
there appears to be no discernible benefit to using higher values of
$s$, while using $s>10$ significantly increases the time for model
computation. For $s=10$, the fits converged on $\mbh= 6.0\times10^8$
\msun, with $\sigmaturb = 24.7$ \kms\ and $\Upsilon_R = 7.25$.

We also ran a model fit in which \chisq\ was computed using the
full-resolution ALMA data cube and model, rather than the $4\times4$
pixel block-averaged version.  With $s=10$, the results for the
best-fitting parameters were virtually identical to the block-averaged
fits: we found best-fit values of $\mbh = 6.0\times10^8$ \msun,
$\sigmaturb = 24.3$ \kms, and $\Upsilon_R = 7.26$. In all subsequent
models fitted to the data cube, we apply the $4\times4$ block
averaging to the data and models when calculating \chisq\ in order to
alleviate any possible issues related to correlated errors in the
background noise, but the block averaging does not appear to have an
appreciable impact on the best-fitting parameter values.

\subsection{Model fits with flat \sigmaturb}
\label{sec:flatsigma}

\begin{figure}
  \scalebox{0.5}{\includegraphics{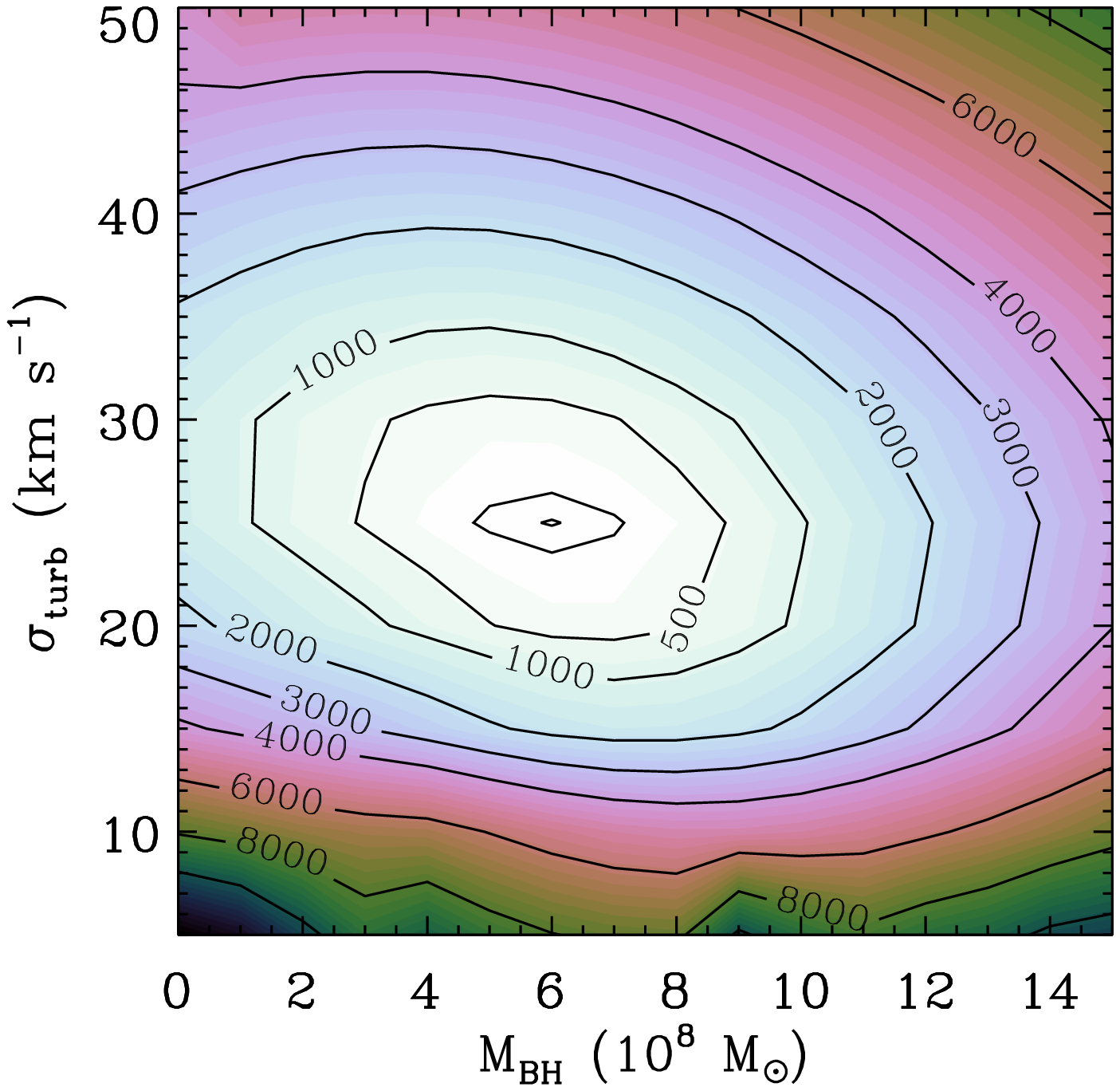}}
  \caption{Contours of constant $\Delta\chisq$ as a function of
    \mbh\ and \sigmaturb, for the flat \sigmaturb\ profile. The
    minimum \chisq\ value of 11173.4 (for 2540 degrees of freedom) is
    obtained at $\mbh=6\times10^8$ \msun\ and $\sigmaturb = 25$
    \kms. The two innermost contours correspond to $\Delta\chisq = 10$
    and 100. Models were calculated over a grid with increments of
    $10^8$ \msun\ and 5 \kms.}
  \label{flatsigma2dgrid}
\end{figure}

With oversampling fixed to $s=10$, we ran initial model optimizations
for a two-dimensional grid over a range of fixed values of \mbh\ (from
0 to $2\times10^9$ \msun\ in increments of $10^8$ \msun) and
\sigmaturb\ (from 5 to 50 \kms\ in increments of 5 \kms). All other
parameters were allowed to vary freely.  Figure \ref{flatsigma2dgrid}
shows contours of constant $\Delta\chisq$ relative to the best-fitting
model at $\mbh=6.0\times10^8$ \msun\ and $\sigmaturb = 25$ \kms, where
the minimum \chisq\ was 11173.4. The value of \chisq\ climbs very steeply
as the parameters depart from these best-fitting values.

For a closer examination of the parameter space around the best fit,
we then ran a grid of models with finer sampling in \mbh\ ranging
from 0 to $2\times10^9$ \msun\ in which \sigmaturb\ and other
parameters were left free. We found that the disk orientation
parameters $i$ and $\Gamma$ converged to very narrow ranges that were
insensitive to the fixed \mbh\ values in these fits, with inclination
$i = 83\arcdeg-84\arcdeg$, and kinematic PA $\Gamma = 117\fdg44 -
117\fdg67$, consistent with the kinematic PA determined by
kinemetry for the outer disk. The disk's centroid velocity and
position parameters also converged tightly, with $\vsys = 1559\pm1$
\kms, and the best-fitting $x$ and $y$ positions for the disk's
dynamical center remaining constant to within $\pm0.1$ ALMA pixels
over the range of \mbh\ values tested.

\begin{figure*}
  \scalebox{0.9}{\includegraphics{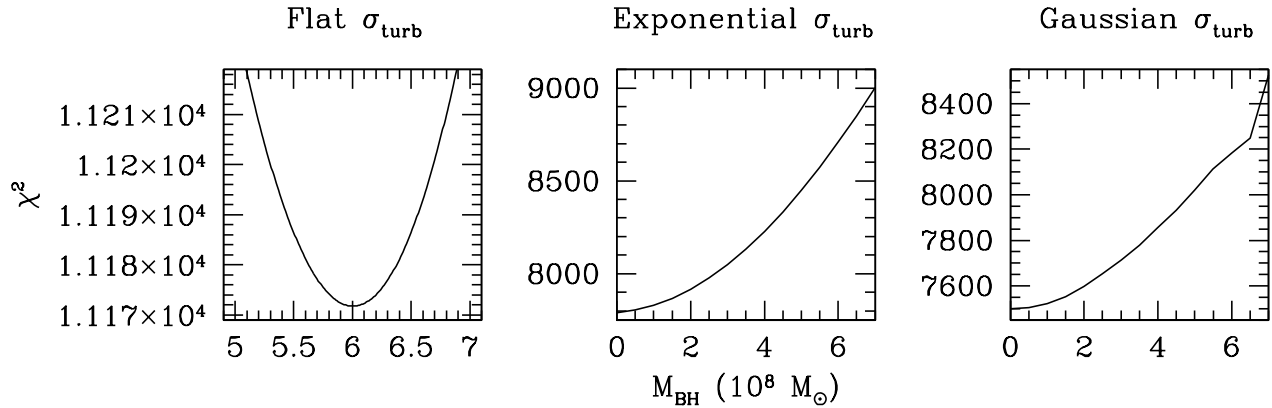}}
  \caption{Curves of \chisq\ as a function of \mbh\ for models
    calculated with the flat, exponential, and Gaussian
    \sigmaturb\ profiles.}
  \label{chisqcurves}
\end{figure*}

For these model fits, the best-fitting value for \sigmaturb\ ranged
from 23 to 27 \kms, slightly larger than one velocity channel in the
ALMA data cube.  The $R$-band mass-to-light ratio is 
anticorrelated with \mbh, and for \mbh\ ranging from 0 to
$2\times10^9$ \msun, $\Upsilon_R$ ranged from 8.10 to 5.31.

The best-fitting model with flat \sigmaturb\ is found at $\mbh =
6.02\times10^8$ \msun.  In this model fit, $\Upsilon_R = 7.25$ and
$\sigmaturb = 24.7$ \kms. With 2540 degrees of freedom, the model has
$\chisq = 11171.8$ and $\chisqdof = 4.40$, indicating a poor fit to
the data overall.  Figure \ref{chisqcurves} shows \chisq\ as a
function of \mbh\ for these model fits. If we were to adopt the usual
criterion of $\Delta\chisq = 6.63$ corresponding to a 99\% confidence
range, we would obtain an extremely narrow range of $(5.66 - 6.34)
\times10^8$ corresponding to an uncertainty of $\pm6\%$ about the
best-fitting value of \mbh. However, since \chisqdof\ is much greater
than unity for the best fit, standard $\Delta\chisq$ intervals are not
directly applicable and we do not consider these confidence intervals
to be meaningful. If we were to inflate the noise uncertainties by a
factor of 2 in order to achieve $\chisqdof \approx 1$ for the best
fit, the $\Delta\chisq = 6.63$ interval would correspond to a slightly
broader mass range of $(5.3-6.7)\times10^8$ \msun.

Figure \ref{momentmaps} illustrates the kinematic moment maps for this
model fit, which reproduce the overall features of the observed moment
maps reasonably well although there are clearly systematic differences
in detail. The model $h_3$ map approximately matches the sign reversal
in $h_3$ seen on either side of the nucleus: the line profiles at
small radii have very extended high-velocity tails, while at large
radii along the disk major axis the line profiles have extended
low-velocity tails. Since the modeled line profiles are assumed to be
intrinsically Gaussian at each location in the disk before beam
smearing, the departures from Gaussian profiles in the modeled data
cube can only result from blending of line profiles originating from
different locations in the disk.  The close match between the observed
and modeled moment maps confirms that the complex line profile shapes
in the data are predominantly the result of rotational broadening and
beam smearing.

\begin{figure}
  \scalebox{0.8}{\includegraphics{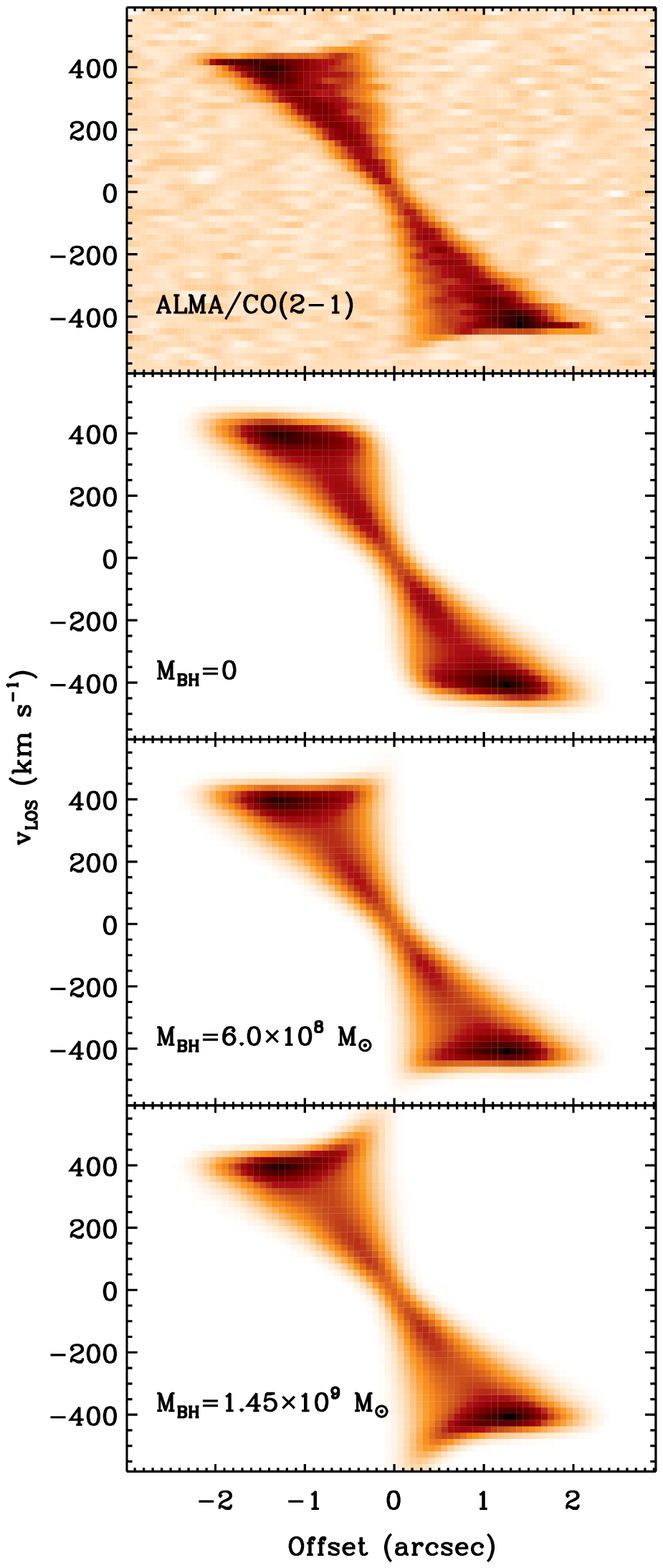}}
  \caption{PVDs for models with flat \sigmaturb\ profiles compared with
    the ALMA data. The three models displayed are the best-fitting
    models with flat \sigmaturb\ for \mbh\ fixed to 0, $6.0\times10^8$,
    and $1.45\times10^9$ \msun. The best-fitting \sigmaturb\ values
    are 27.7, 24.7, and 24.0 \kms, respectively.}
  \label{pvdfig-flatsigma}
\end{figure}

Figure \ref{pvdfig-flatsigma} shows the best-fitting model PVDs for
models calculated with \mbh\ fixed to 0, $6.0\times10^8$, and
$1.45\times10^9$ \msun; the last value is equal to the best-fitting
mass from \citet{rusli2011}.  The model with $\mbh=0$ clearly fails in
that it does not have a central upturn in maximum rotation
velocity. At $\mbh=6\times10^8$ \msun, the model has a central
velocity upturn similar to that seen in the data, and at
$\mbh=1.45\times10^9$ \msun\ the central upturn is  too
prominent to match the data.

\begin{figure*}
  \begin{center}
    \scalebox{0.6}{\includegraphics{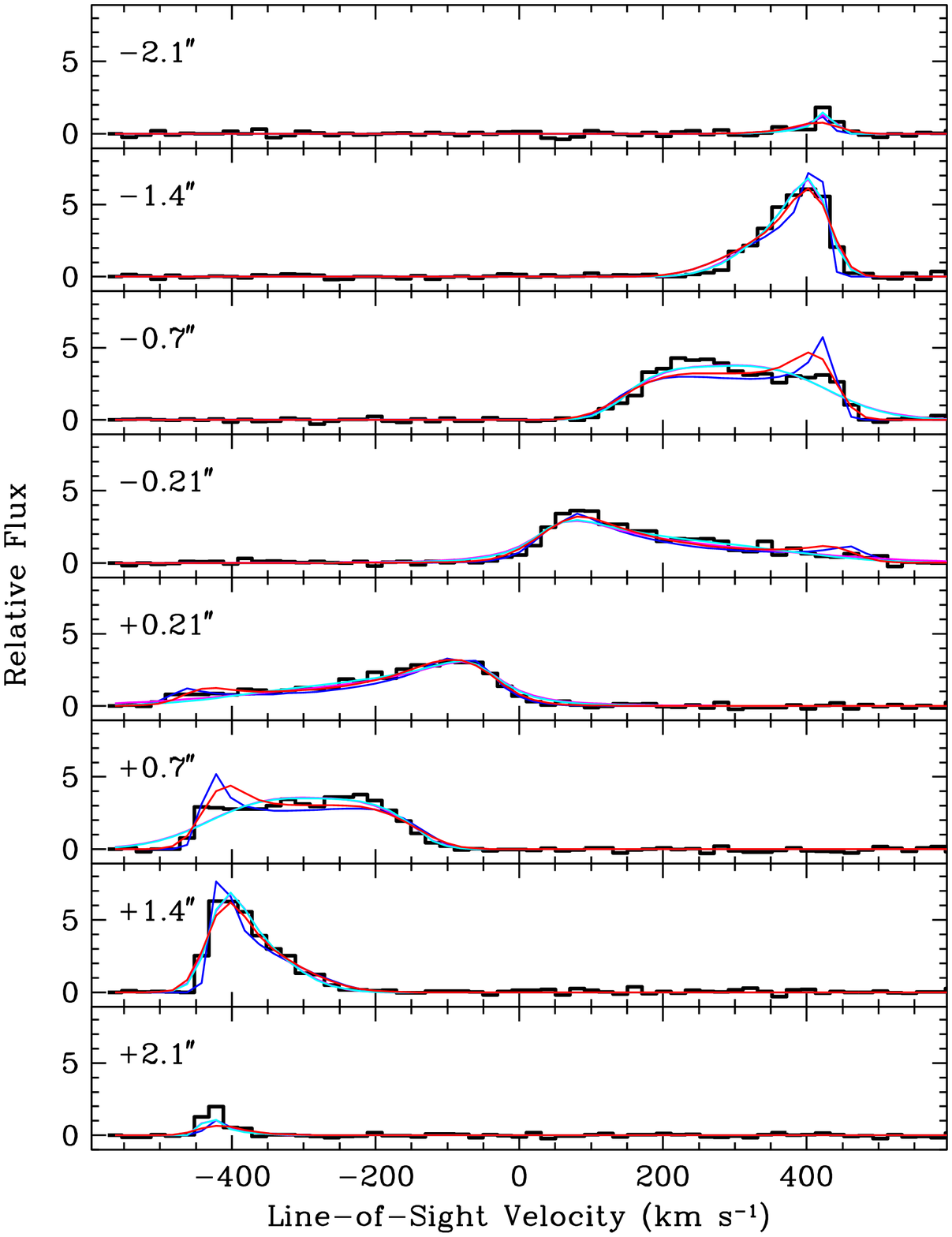}}
    \end{center}
  \caption{Line profiles measured by extracting single-column cuts
    through the PVD at eight spatial locations. The PVD is based on a
    four pixel wide extraction along the disk major axis. The observed
    PVD is shown in black. Modeled profiles are as
    follows. \emph{Red:} The best-fitting model with flat
    \sigmaturb. \emph{Blue:} Model with flat $\sigmaturb=10$
    \kms. \emph{Magenta:} Best-fitting model with exponential
    \sigmaturb. \emph{Cyan:} Best-fitting model with Gaussian
    \sigmaturb. The models with exponential and Gaussian
    \sigmaturb\ profiles are closely overlapping at most locations.
    Numerical labels in each panel show the offset in arcseconds from
    the galaxy nucleus along the major axis. }
  \label{pvdslices}
\end{figure*}

One noticeable aspect of these model PVDs is that the CO linewidths in
the outer disk are broader than in the data. This can also be seen in
Figure \ref{pvdslices}, which illustrates line profiles extracted from
the PVD at eight spatial locations.  The models follow the overall
shapes of the observed line profiles fairly well, even near the disk
center at $\pm0\farcs21$ (or $\pm3$ pixels) where the profiles are
extremely broad and asymmetric. In the outermost slices through the
PVD, the modeled CO profiles are clearly broader than the observed
profiles.

\begin{figure}
  \scalebox{0.85}{\includegraphics{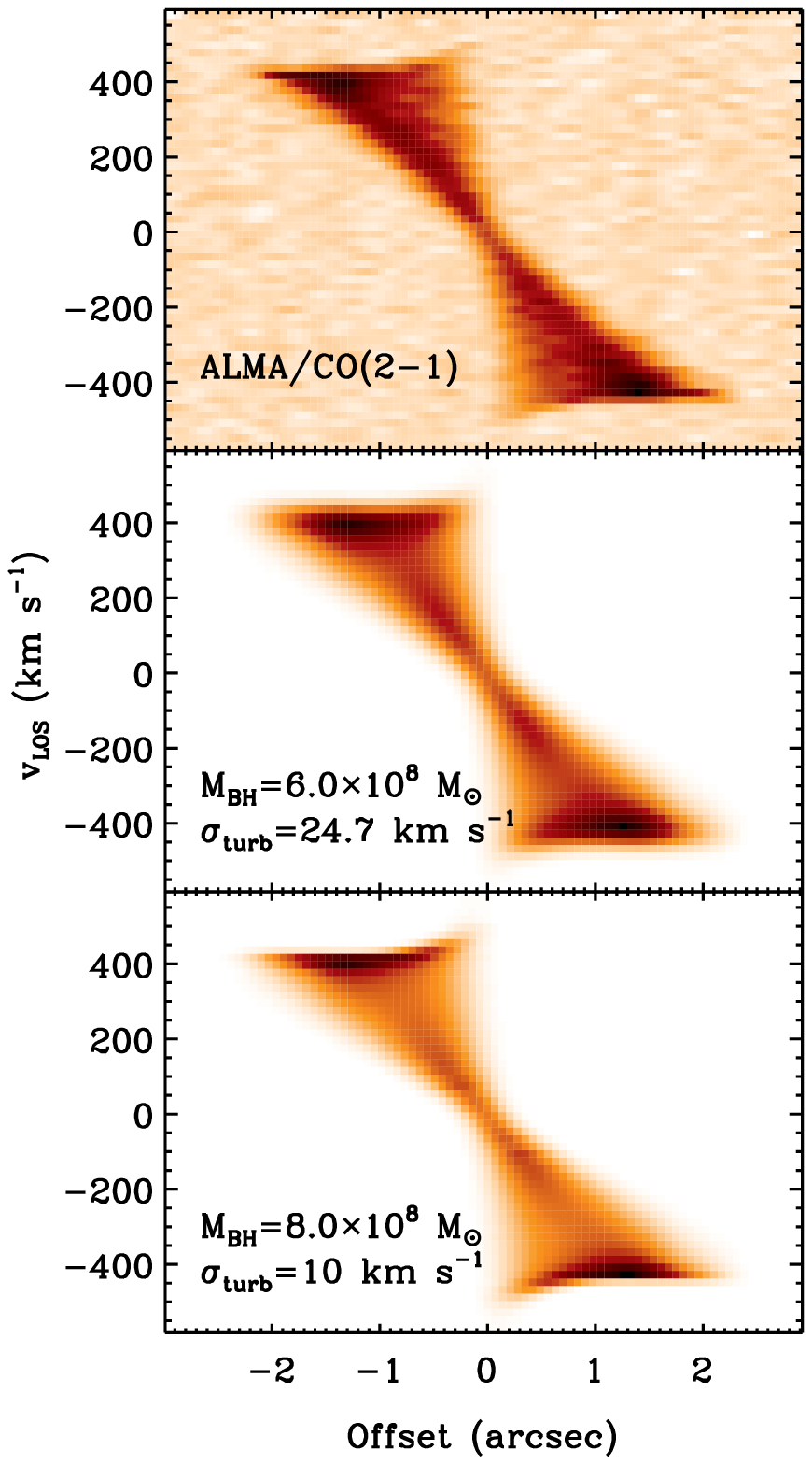}}
  \caption{PVDs for models with flat \sigmaturb\ profiles. The middle
    panel shows the overall best-fitting model with flat \sigmaturb,
    having $\mbh=6.0\times10^8$ \msun\ and $\sigmaturb = 24.7$ \kms.
    The lower panel shows the best fitting model having a fixed
    $\sigmaturb=10$ \kms. For this model, $\mbh=8.0\times10^8$ \msun.}
  \label{pvdfig-flatsigma2}
\end{figure}

To examine the possibility that the model optimization is somehow
biased towards high values of \sigmaturb, we created a PVD for the
best-fitting model having $\sigmaturb=10$ \kms; at this value, the
intrinsic CO linewidths would be unresolved by the ALMA data. For
$\sigmaturb=10$ \kms, the best-fitting BH mass is $8.0\times10^8$
\msun.  The value of \chisq\ for this model is 16063.5, dramatically
worse than the best-fitting model at $\mbh=6.0\times10^8$ \msun\ (as
can be seen in Figure \ref{flatsigma2dgrid}). Figure
\ref{pvdfig-flatsigma2} compares the PVD of this model with the model
having $\mbh=6.0\times10^8$ \msun\ and $\sigmaturb=24.7$ \kms. It is
somewhat striking that the $\sigmaturb=10$ \kms\ model appears to
match some aspects of the observed PVD distinctly better than does the
best-fitting model. In particular, the $\sigmaturb=10$ \kms\ model
appears to more closely match the observed narrow linewidths in the
outer disk, and the shape of the inner  velocity upturn at
small radii. The model is, however, a much worse match to the observed
PVD overall, as clearly indicated by its much larger
\chisq\ value. The line-profile cuts through the modeled PVD, shown in
Figure \ref{pvdslices}, help to clarify how this model fails to match
the data. At positions of $\pm0\farcs7$ from the disk center, there is
a ``pile-up'' of emission at $\vlos = \pm420$ \kms\ causing a spike in
emission that is not present in the data at these locations. Larger
values of \sigmaturb\ in the inner regions of the disk tend to reduce
the amplitude of this emission peak, matching the data more closely.

\subsection{Model fits with other \sigmaturb\ profiles}

Since the best-fitting model with $\sigmaturb=24.7$ \kms\ clearly
overpredicts the CO linewidths in the outer disk, while models with
smaller values of $\sigmaturb$ lead to dramatically poorer fits to the
data overall, we conclude that the flat $\sigmaturb$ profile is not a
sufficient description of the disk's turbulent velocity dispersion
profile. To allow for radial gradients in \sigmaturb\ and attempt to
achieve more satisfactory fits with lower \chisq, we ran grids of
models using the exponential and Gaussian \sigmaturb\ profiles. As
before, models were run over a set of fixed values of \mbh, with all
other parameters allowed to vary freely. 

Figure \ref{chisqcurves} shows \chisq\ as a function of
\mbh\ for both the exponential and Gaussian \sigmaturb\ profiles.  The
most basic result is that the more complex \sigmaturb\ profiles lead
to better fits overall with significantly lower \chisq, but taken at
face value they imply a BH mass much lower than the value derived from
the flat \sigmaturb\ fits.

For the exponential \sigmaturb\ profile, when all parameters were
allowed to vary freely, the model fits converged on a best-fitting BH
mass of zero (Figure \ref{chisqcurves}, middle panel), with
$\Upsilon_R=7.94$. For this best-fitting model ($\chisq=7790.8$ for
2539 degrees of freedom) the central value of \sigmaturb\ becomes
quite high, with $\sigma_0=268$ \kms\ and $r_0 = 60.7$ pc. The PVD for
this model, shown in Figure \ref{pvdfig-sigmaprofiles}, helps to
illustrate why this model produces a much lower \chisq\ than models
with flat \sigmaturb. At radii between 1\arcsec\ and 2\arcsec, the PVD
of the exponential \sigmaturb\ model is a much better match to the
observed PVD structure than any of the flat \sigmaturb\ models. At the
same time, the fit with $\mbh=0$ does not have a central
velocity upturn. Instead, the smallest radii in the PVD are
characterized by an increase in linewidth due to the high central
\sigmaturb\ value. In effect, the model fit is unable to distinguish
between rotation and dispersion in the inner disk, while in the outer
disk the fits clearly prefer having a gradient in \sigmaturb\ allowed
by the exponential model, as opposed to the flat
\sigmaturb\ models. Qualitatively, the structure of the PVD is a very
poor match to the central velocity upturn of the data, but
the \chisq\ minimization result is dominated by structure at larger
radii in the PVD.

\begin{figure}
  \scalebox{0.45}{\includegraphics{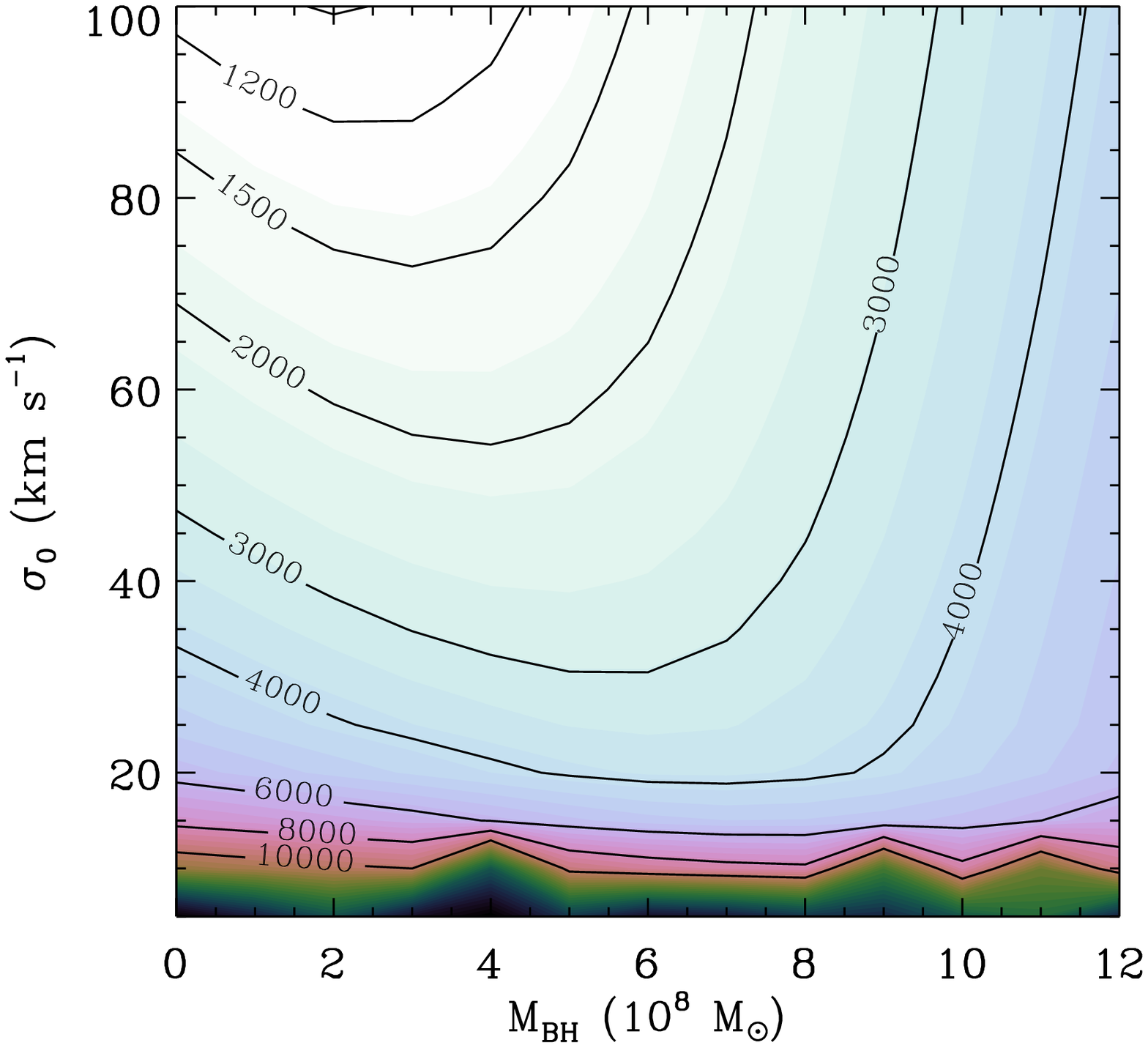}}
  \caption{Contours of constant $\Delta\chisq$ for models with the
    exponential \sigmaturb\ profile, as a function of \mbh\ and the
    central turbulent velocity dispersion $\sigma_0$.  For
      $\sigma_0 \le 50$ \kms, the best-fitting BH mass is
      $\sim(4-8)\times10^8$ \msun, but higher fixed values of
    $\sigma_0$ result in model fits with lower \mbh. If $\sigma_0$ is
    left as a free parameter, the best-fitting model has
    $\sigma_0=268$ \kms, and $\chisq = 7790.8$ for 2539 degrees of
    freedom; the $\Delta\chisq$ values plotted in this figure are
    relative to that best-fitting value.  }
  \label{expsigma-2dgrid}
\end{figure}

\begin{figure}
  \scalebox{0.85}{\includegraphics{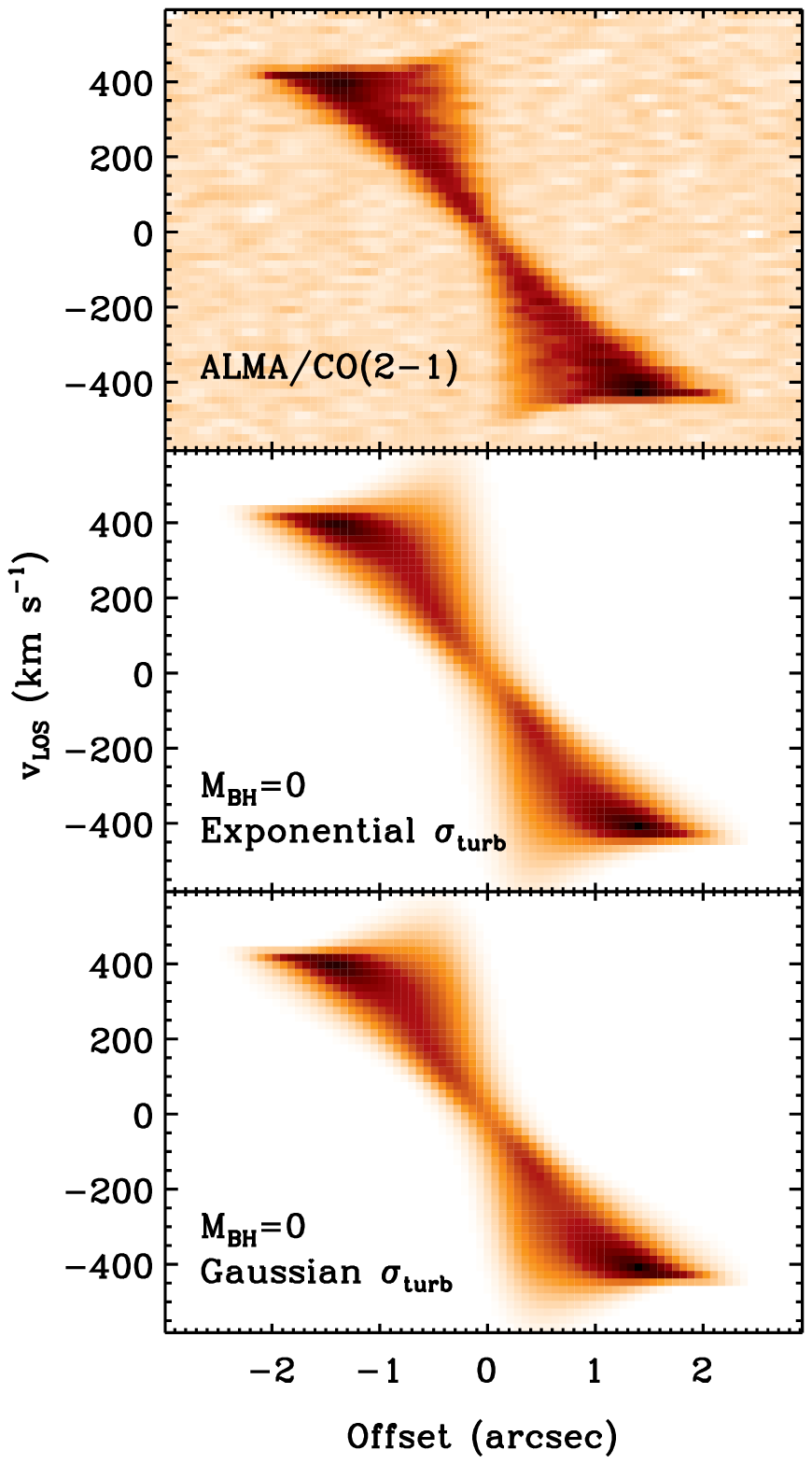}}
  \caption{PVDs for models with the exponential and Gaussian
    \sigmaturb\ profiles. Compared with PVDs for the flat
    \sigmaturb\ model (Figure \ref{pvdfig-flatsigma2}), these PVDs fit
    the data much better at large radii in the disk ($r>1\arcsec$) and
    give much lower \chisq\ values, but are a poor match to the
    central velocity upturn.}
  \label{pvdfig-sigmaprofiles}
\end{figure}

While a central turbulent velocity dispersion of 268 \kms\ is probably
unphysically high for molecular gas, it is not clear whether any
specific prior constraint on $\sigma_0$ would be justifiable. To
examine the interplay between \mbh\ and constraints imposed on the
central value of \sigmaturb, we ran a grid of models over a range of
fixed values of \mbh\ and of the central turbulent velocity dispersion
$\sigma_0$. The results are shown in Figure \ref{expsigma-2dgrid}. We
find that when $\sigma_0$ is restricted to low values in the range
$\sim10-30$ \kms, similar to the range of turbulent velocity
dispersions seen in resolved GMCs in nearby galaxies, the model fits
converge on $\mbh$ in the range $\sim(6-8)\times10^8$ \msun.  However,
higher fixed values of $\sigma_0$ produce fits with significantly
lower BH masses. Additionally, restricting $\sigma_0$ to low values
produces much worse fits overall in comparison with the free
$\sigma_0$ fits: for example, when $\sigma_0$ is fixed to 25 \kms, the
best fit is found for $\mbh=6.0\times10^8$, with $\chisq=11083.0$.

This illustrates the most severe problem in fitting models to the NGC
1332 data: once we allow for the possibility of a radial gradient in
\sigmaturb, the inferred value of \mbh\ is directly dependent upon the
assumed upper limit to \sigmaturb\ in the central region of the
disk. There does not appear to be any clear way to circumvent this
problem, since the line profiles are so strongly affected by
rotational broadening that \sigmaturb\ cannot be directly measured or
constrained independently of a dynamical model for the disk
rotation. Larger values of \sigmaturb\ directly lead to significantly
lower \chisq\ for the overall model fit, and by the \chisq\ criterion
the model fits strongly prefer a model with no BH but with a central
\sigmaturb\ that is probably unphysically large. It is also important
to note that for the highest values of \sigmaturb\ reached in these
models, the thin disk assumption would break down, invalidating the
basic premises of our dynamical modeling. If the disk were actually
highly turbulent in its central regions, the turbulent pressure
support would have to be accounted for in determining \mbh. This would
raise \mbh\ to a non-zero value.

We also ran a separate grid of models with the exponential
\sigmaturb\ profile and $s=25$, to test whether the high central
\sigmaturb\ values might be the result of insufficient model
resolution of the disk's central kinematics at $s=10$. The model
fitting results were essentially identical to the $s=10$ case.

As an attempt to examine a model in which \sigmaturb\ possesses a
radial gradient without a central peak, we ran model fits using the
Gaussian \sigmaturb\ profile. Since the Gaussian centroid can be
radially offset from the disk center, this model allows for a central
depression in \sigmaturb. For a grid of model fits over fixed values
of \mbh, Figure \ref{chisqcurves} (right panel) shows the best-fitting
\chisq\ as a function of \mbh.  Similar to the exponential profile
models, we find that the best fit with the Gaussian
\sigmaturb\ profile is obtained for $\mbh=0$ and $\Upsilon_R=7.96$,
and the best-fitting \chisq\ value in this case is 7499.7 for 2538
degrees of freedom. In this model, the Gaussian \sigmaturb\ parameters
are $\sigma_0 = 100.3$ \kms, $\mu = 78.9$ pc, and $r_0 = 14.5$
pc. This small value of $r_0$ indicates that the Gaussian peak is very
close to the disk center, and the central dip in \sigmaturb\ is
essentially unresolved. Trial fits demonstrated that if $\sigma_0$ is
restricted to values well below 100 \kms, the inferred \mbh\ value
from \chisq\ minimization is directly determined by the constraint
imposed on $\sigma_0$, similar to the situation for the exponential
\sigmaturb\ profile. The PVD for this model is displayed in Figure
\ref{pvdfig-sigmaprofiles}, and Figure \ref{pvdslices} shows the CO
line profiles at several locations along the disk major axis for the
best-fitting exponential and Gaussian \sigmaturb\ models.

The best-fitting models with the exponential or Gaussian
\sigmaturb\ profiles also converged on values of $i$ between
83\arcdeg\ and 84\arcdeg, and $\Gamma$ between 117\arcdeg\ and
118\arcdeg, consistent with results from the flat \sigmaturb\ models.

\subsection{Stellar Mass Profile}

As a test to assess the sensitivity of \mbh\ to the inner mass profile
of the galaxy, we ran a model fit using the mass profile measured from
the \hst\ F814W image with $\ml$ as a free parameter, using the flat
\sigmaturb\ profile. The model converged on a best-fitting mass of
$\mbh = 9.85\times10^8$ \msun\ and $\sigmaturb=25.0$ \kms, and the
stellar mass profile for the best fit is illustrated in Figure
\ref{massprofiles}. The best fit has $\chisq = 11690.8$, compared with
the value 11171.8 for the best fit using the \citet{rusli2011} mass
profile and flat \sigmaturb.

It is reasonable to assume that the stellar luminosity profile
measured from $K$-band data is a more accurate representation of the
galaxy structure, having less sensitivity to dust extinction, and the
$K$-band galaxy profile gives a lower \chisq\ value for the dynamical
model fit. The difference in BH masses measured with the two profiles
illustrates the importance of using a mass profile measured from
near-IR imaging whenever possible, to minimize the impact of
extinction.  Higher-resolution kinematic data would also lessen the
impact of extinction on the derived \mbh; when the central, nearly
Keplerian region of the disk is better resolved, the impact of errors
in $M_\star(r)$ on the derived \mbh\ will become smaller.

\subsection{Fits to the PVD}

Another option for model fitting is to carry out fits to the PVD
rather than to the full data cube. This may have an advantage in that
the PVD describes the major axis kinematics of the disk, and fitting
models to the PVD rather than to the full data cube could have better
sensitivity to the central velocity upturn and lead to
better constraints on \mbh. However, fits to the PVD would not be able
to constrain the disk inclination, kinematic PA, or centroid position
as well as fits to the data cube, and these parameters would need to
be constrained or fixed.

We carried out a fit to the PVD using the flat \sigmaturb\ model, with
the values of $i$, $\Gamma$, and the $x$ and $y$ centroid positions
fixed to their best-fitting values from the corresponding fit to the
full data cube. Free parameters included \mbh, \ml, \sigmaturb, \vsys,
and \fscale.  For each model iteration, we extracted a PVD
following the same procedures used on the data, with a four pixel
extraction width. For the calculation of \chisq, we measured the
background noise as the standard deviation of pixel values in blank
regions of the PVD. We note that the PVD still exhibits strongly
correlated noise among pixels along the position axis over a scale of
$\sim4$ pixels.

This model run gave best-fitting values of $\mbh=6.5\times10^8$ \msun,
$\Upsilon_R=7.18$, and $\sigmaturb=21.6$ \kms.  Thus, in this case
fitting to the PVD rather than to the data cube altered the
best-fitting value of \mbh\ by just 8\%. This can be attributed to the
fact that the disk is so highly inclined that a 4-pixel wide
extraction of the PVD already contains much of the kinematic
information of the disk as a whole.

The fits to the full data cube demonstrated that \chisq\ minimization
has a tendency to optimize the model fit to the outer disk, with
relatively low sensitivity to the small number of pixels in the
central velocity upturn region. We therefore carried out
another model run in which \chisq\ was calculated only over a
restricted set of pixels in the PVD corresponding to the spatial and
velocity region sensitive to the central upturn. For this
model run, we fixed \ml\ to its best-fitting value from the previous
fit to the full PVD, leaving \mbh, \sigmaturb, \vsys, and
\fscale\ as free parameters. The calculation of \chisq\ was
carried out only for pixels in the PVD corresponding to $r<1\arcsec$
and $\vlos > 300$ \kms.  This model run gave $\mbh=5.7\times10^8$
\msun\ and $\sigmaturb=37.7$ \kms.  As in the case of fitting to the
full PVD, fitting models with $\sigmaturb$ restricted to lower values
led to higher values of \mbh\ (but higher \chisq): enforcing an upper
limit of $\sigmaturb=10$ \kms\ gave $\mbh = 8.0\times10^8$
\msun. Since the best-fitting \mbh\ values were very similar to the
values derived from fits to the full data cube, we conclude that for
this dataset there is no clear advantage to fitting models to either
the full PVD or to the central high-velocity region of the PVD.

\subsection{Final Results}

The dynamical model fits do not lead to a definitive determination of
\mbh, primarily because the inferred value of \mbh\ depends very
directly on the specific choice of some parameterized model for
$\sigmaturb(r)$ or on any upper limit imposed on the value of
$\sigmaturb(r)$.  The simplest model, with the flat
\sigmaturb\ profile, converges tightly to $\mbh=6.0\times10^8$
\msun. The best-fitting model has $\chisqdof = 4.4$, however, and the
line profiles shown in Figure \ref{pvdslices} demonstrate that
\chisq\ is dominated by localized regions where the modeled profiles
deviate systematically from the data, indicating that the models are
failing to capture some essential structure in the disk's velocity
field, turbulent velocity dispersion profile, or surface brightness
profile. In the outer disk, the observed line profiles are clearly
narrower than the best-fitting overall value of $\sigmaturb = 24.7$
\kms, but when \sigmaturb\ is fixed to lower values, the line profiles
at intermediate radii in the disk deviate even more strongly from the
data (also visible in Figure \ref{pvdslices}). The central
high-velocity envelope of the PVD appears qualitatively to be better
matched by a model with $\sigmaturb$ fixed at 10 \kms\ (implying $\mbh
= 8\times10^8$ \msun) than with $\sigmaturb = 24.7$ \kms, but the
model with $\sigmaturb = 10$ \kms\ is a much worse fit overall in
terms of \chisq. Restricting the model fit to a small sub-region of
the PVD at small radius and high velocity does not alter these results
substantially, compared with fits to the full data cube. The
systematic deviations between models and data can also be seen in the
kinematic moment maps (Figure \ref{momentmaps}). While the models
reproduce the overall structure of the kinematic maps reasonably, even
up to $h_5$ and $h_6$, there are obvious differences in detail, which
can plausibly be attributed to an inadequate model for the spatial
distribution of \sigmaturb\ in the disk.

Models in which \sigmaturb\ is allowed to vary with radius give much
lower values of \chisqdof, although still well above unity. Fits
carried out with the exponential and Gaussian \sigmaturb\ profiles
drive \mbh\ to zero, substituting dispersion for rapid rotation in the
inner region of the disk. These models are formally more successful
than the flat \sigmaturb\ models because they fit the outer disk much
better while matching the inner high-velocity envelope of the PVD
relatively poorly. The lowest \chisq\ value is found for the Gaussian
\sigmaturb\ model fit, which gives $\chisqdof = 2.95$ at $\mbh=0$. The
appearance of the PVDs for the exponential and Gaussian
\sigmaturb\ models gives a fairly clear indication that the high
central dispersions are spurious, however. These fits imply central
values of \sigmaturb\ that are probably unphysically large (over 250
\kms\ in the case of the exponential profile), but restricting the
maximum \sigmaturb\ amplitude to lower values produces significantly
worse fits with much higher \chisq\ (Figure \ref{expsigma-2dgrid}).
With the exponential profile, a limit on the maximum value of
\sigmaturb\ ($<50$ \kms) based on observations of resolved GMCs in
other environments leads to \mbh\ in the range $(4-8)\times10^8$
\msun.

These difficulties in constraining \mbh\ appear to be driven by a
combination of factors: the models systematically deviate from the
data in some locations of the disk by amounts much larger than the
observational uncertainties; the beam-smeared line profiles near the
disk center are so broad that the model fits are unable to distinguish
clearly between rotation and turbulence as the source of the
broadening; a relatively small number of data pixels are located in
spatial and velocity regions of the data cube that are highly
sensitive to \mbh; and the S/N in those \mbh-sensitive pixels is
relatively low.

Fortunately, the disk inclination and orientation parameters, centroid
position, and systemic velocity are well constrained by the model
fits, and we obtain consistent values for these parameters regardless
of the \sigmaturb\ prescription that is used. The stellar
mass-to-light ratio in the best-fitting models is consistent with the
value of $\ml_R=7.35$ from \citet{rusli2011}. For the models with
flat, exponential, and Gaussian \sigmaturb, we obtain best-fitting
values of $\ml_R = 7.25$, 7.94, and 7.96, respectively. The two latter
cases should be considered effectively as upper limits on $\ml_R$,
however, since these model fits converged on $\mbh=0$.

The flat and exponential \sigmaturb\ models together point to a BH
mass that is likely to be in the range $\sim(4-8)\times10^8$ \msun,
under the assumption that a plausible maximum value for \sigmaturb\ in
the inner disk is $\sim50$ \kms. We adopt this as a provisional and
preliminary conclusion, but we emphasize that this is not a
quantitatively rigorous result.  We are unable to derive meaningful
confidence limits on \mbh\ owing to the high \chisq\ values for our
models as well as the fact that the best \chisq\ values (by far) are
found for models with extremely large (and probably spurious) values
of \sigmaturb\ in the inner portion of the disk.  With no clear way to
determine $\sigmaturb(r)$ independently or to constrain its maximum
possible value, a firm lower limit on \mbh\ cannot be derived from the
Cycle 2 data. In all of our models, \chisq\ rises steeply at
$\mbh>10^9$ \msun; the disk kinematics do not appear to be compatible
with the value $\mbh=(1.45\pm0.20)\times10^9$ \msun\ found by
\citet{rusli2011}.

\section{Discussion}

The 0\farcs3 resolution ALMA observation of NGC 1332 permits an
instructive case study for gas-dynamical BH detection. With this
dataset, we are working in the regime where the BH sphere of influence
may be marginally resolved, in that the central velocity upturn is
visible as the upper envelope to the PVD at small radii, but the
central high-velocity rotation is badly blurred with low-velocity
emission due to beam smearing and the disk's nearly edge-on
inclination. In essence, \rg\ appears to be nearly resolved along the
disk major axis, but it is badly unresolved along the disk's minor
axis. This makes it difficult to derive strong constraints on
\mbh\ from the information contained in the central velocity upturn.
The difficulties in modeling the NGC 1332 disk dynamics described here
are particularly acute since the disk is very close to edge-on, but we
anticipate that these same issues will arise in any situation in which
the BH sphere of influence is not well resolved.

\subsection{The BH sphere of influence in gas-dynamical measurements}

The ``radius of influence'' for BHs is generally defined as the radius
within which the BH is the dominant contribution to the galaxy's mass
profile. In the absence of a mass model for the galaxy, \rg\ is taken
to be the radius within which the circular velocity due to the
gravitational potential of the BH rises above the surrounding bulge
velocity dispersion \sigmastar: $\rg = GM/\sigmastar^2$.  Since
\sigmastar\ is not spatially constant within a galaxy bulge, and its
central value is affected by the BH itself, this definition does not
give a uniquely well-defined value of \rg, but it does provide a
useful general guideline for the radius that should be resolved in
order to detect the dynamical effect of the BH. The gravitational
influence of the BH can be detected at radii beyond \rg, but at
progressively larger radii the enclosed mass becomes dominated by
stars. Measuring \mbh\ when \rg\ is unresolved then becomes an
exercise in detecting a small fractional contribution to the total
gravitating mass on unresolved spatial scales, and the results can be
highly susceptible to systematic errors in determination of the
stellar mass profile or in other aspects of the model construction.

The criterion of resolving \rg\ as the minimal requirement for clear
detection of the dynamical effect of the BH ignores one crucial
factor, highlighted by the NGC 1332 data. Along the minor axis of an
inclined disk, the projected distance from the nucleus to a point at
distance \rg\ from the nucleus is compressed by a factor of $\cos
i$. This is an important effect for highly inclined disks: for NGC
1332, $\cos i \approx 0.11$. In an observation just sufficient to
resolve \rg\ along the disk major axis in NGC 1332, \rg\ along the
disk's projected minor axis would be unresolved by an order of
magnitude. In the ALMA data described here, the poor spatial
resolution along the disk's minor axis direction leads directly to
severe beam smearing, causing the line profiles to be dominated by
low-velocity emission even at small radii along the disk major axis.

Our examination of the NGC 1332 disk dynamics suggests that for
gas-dynamical BH mass measurements, the key criterion for whether the
BH sphere of influence is resolved should be whether \rg\ is resolved
along the disk's projected \emph{minor} axis, not along its major axis
as is usually assumed.  In other words, to ensure that the central
velocity upturn is clearly resolved and not severely spatially blended
with low-velocity emission, the observations should resolve an angular
scale corresponding to $\rg\cos i$. For highly inclined disks such as
NGC 1332, the requirements on angular resolution for dynamical
detection of BHs are thus much more stringent than for disks at
moderate inclination angles. BH mass measurements from disks observed
at very low inclination angles will present a different set of
challenges, and will be particularly difficult if $\sin i$ is so small
that that the line-of-sight component of the disk's rotational
velocity is not much larger than the turbulent or random velocities in
the disk. All else being equal, gas-dynamical BH mass measurements
will require higher velocity resolution for disks of lower
inclination, and higher angular resolution for high inclinations.

To what extent is \rg\ resolved in NGC 1332 with the Cycle 2 ALMA
observation?  Assuming $\mbh=6\times10^8$ \msun\ and using $\sigmastar
= 328$ \kms\ \citep{kormendyho}, the kinematic definition of the
radius of influence gives $r_g = 24$ pc. This is equivalent to an
angular radius of 0\farcs22, indicating that the BH radius of
influence along the disk major axis would be slightly unresolved. In
contrast, $r_g \cos i = 0\farcs027$, so the radius of influence is
unresolved by an order of magnitude in the minor-axis direction. We
can also use the dynamical modeling results to estimate the size of
\rg\ as the radius of the sphere within which $M_\star(r) = \mbh$.
For $\mbh = 6.0\times10^8$ \msun, our best-fitting model with the flat
\sigmaturb\ profile also gives $\rg = 24$ pc, so the two definitions
of \rg\ give identical results in this case. For the model having
$\sigmaturb=10$ \kms\ and $\mbh=8\times10^8$ \msun, \rg\ is slightly
larger at 29.5 pc or 0\farcs27 along the disk major axis, similar to
the 0\farcs27 resolution of the ALMA data. The stellar-dynamical value
of $\mbh=1.45\times10^9$ \msun\ from \citet{rusli2011} implies a
larger sphere of influence, $\rg = 58$ pc or 0\farcs54.

\begin{figure}
  \scalebox{0.4}{\includegraphics{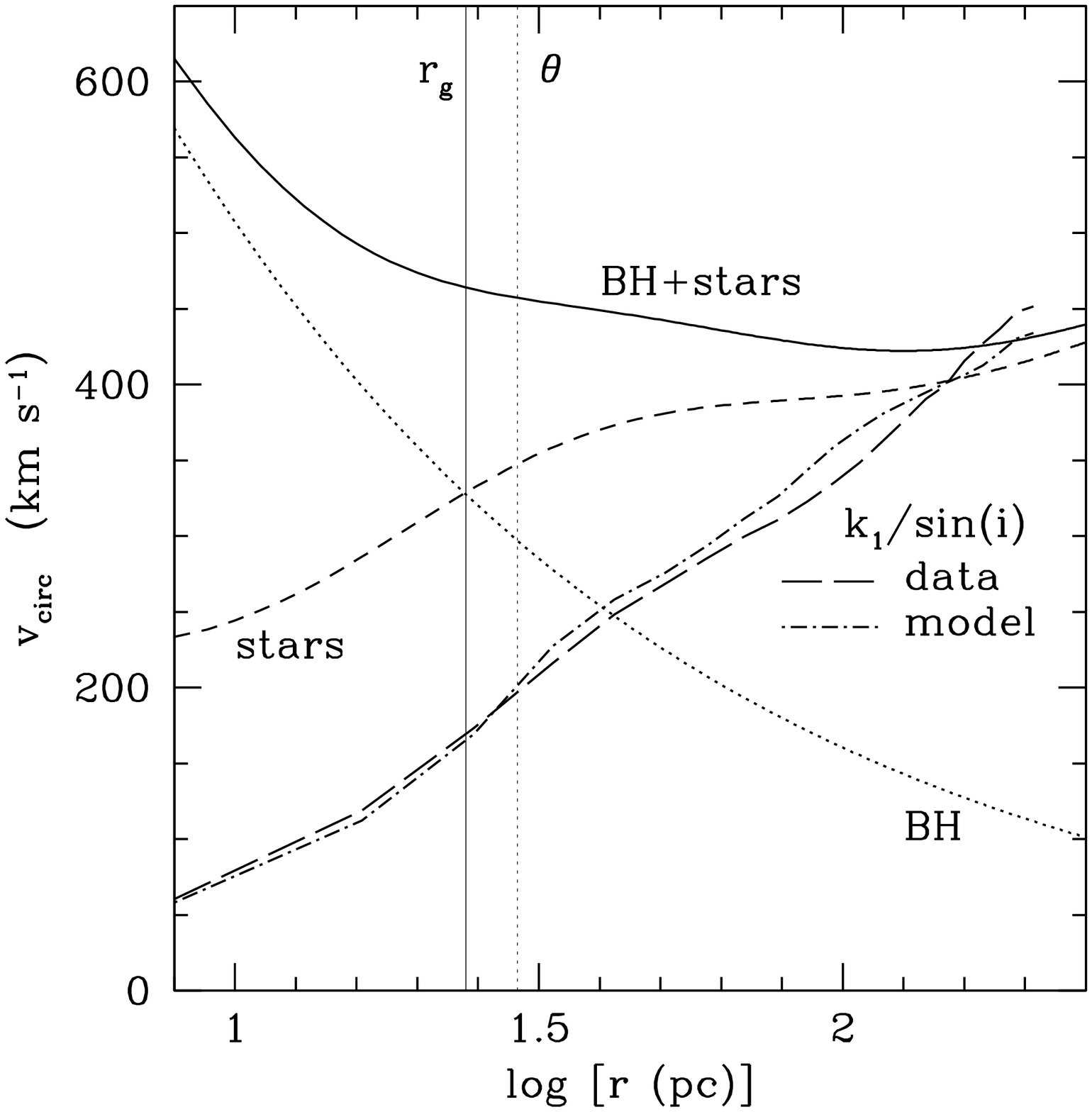}}
  \caption{Plot of circular velocity \vcirc\ as a function of radius,
    to illustrate the impact of beam smearing on the major axis
    velocity profile. Model profiles are calculated for the
    best-fitting model with $\mbh=6.0\times10^8$ \msun\ and
    $\sigmaturb = 24.7$ \kms. \emph{Dotted Curve:} $\vcirc(r)$ for the
    BH alone. \emph{Short-dashed curve:} $\vcirc(r)$ for the host
    galaxy stellar mass alone. \emph{Solid curve:} $\vcirc(r)$ for the
    combined profile of the BH and stars. \emph{Long-dashed curve:}
    The major-axis velocity profile $k_1/\sin(i)$ measured from the
    observed ALMA \vlos\ map using kinemetry, where $\sin(i) = 0.994$.
    \emph{Dot-dashed curve:} Kinemetry profile of $k_1/\sin(i)$
    measured from the \vlos\ map of the model, which closely follows
    that of the data.  The dotted vertical line gives the radius
    corresponding to the resolution limit of the ALMA data ($\theta$),
    and the solid vertical line denotes $r_g$ for $\mbh =
    6.0\times10^8$ \msun. Beam smearing and the disk's high
    inclination result in an observed major-axis velocity curve
    that falls far below the disk's circular velocity at nearly all
    radii in the disk.}
  \label{beamsmear}
\end{figure}

Figure \ref{beamsmear} illustrates the severity of the beam-smearing
effect for the NGC 1332 data. This figure presents the modeled
circular velocity curve $\vcirc(r)$ for the best-fitting model having
flat $\sigmaturb$ and $\mbh=6.0\times10^8$ \kms, including curves
showing $\vcirc(r)$ due to the BH alone, the stars alone, and the
combined gravitational potential of the BH and stars. For comparison
with the data, we take the $k_1(r)$ profile from kinemetry and divide
by $\sin i$ to produce an observed profile of mean rotation velocity
along the disk major axis, and we also display the $k_1(r)$ profile
measured from a kinemetry fit to the \vlos\ map of the dynamical
model. (With $i\approx83\arcdeg$, $\sin i$ is so close to unity that
it can be essentially ignored.)  In the limit of extremely high
angular resolution, the $k_1$ curve would closely follow the curve of
\vcirc\ for the combined mass profile of the BH and stars.  As a
consequence of beam smearing and the high disk inclination, the
observed line-of-sight centroid velocity $k_1$ falls far below the
actual major-axis \vcirc\ profile at nearly all radii in the disk,
even at radii significantly larger than one resolution element from
the galaxy nucleus. One might naively expect that $k_1$ would closely
track \vcirc\ from large radii down to approximately the resolution
limit of the data, but in fact $k_1$ begins to deviate below
\vcirc\ at a radius roughly five times larger than the angular
resolution limit. 

This analysis further reinforces the conclusion that, although \rg\ is
likely to be nearly resolved along the disk's major axis, the mean or
centroid velocities measured at locations along the disk major axis
have essentially no sensitivity to \mbh. When the information about
\mbh\ is contained primarily in the \emph{shapes} of the beam-smeared
line profiles rather than in the centroid velocity \vlos\ measured
from the line profile at each position, it becomes extremely
challenging to derive accurate constraints on \mbh.

For our fiducial mass model displayed in Figure \ref{beamsmear}, the
presence of emission out to $\pm500$ \kms\ from the systemic velocity
(as observed in the PVD) implies that the disk emission extends down
to $r\approx15$ pc, well inside \rg\ (provided that this outermost
velocity is primarily the result of rotation rather than
turbulence). The upper limit to the observed rotation speed could be
due to the presence of a ``hole'' in the molecular disk at $r<15$ pc,
or simply from a central surface brightness at $r<15$ pc too low to
detect in this observation. This scale is well inside the $r\approx29$
pc angular resolution of our data.

\citet{davis2014} proposed a new figure of merit for gas-dynamical BH
mass measurements as an alternative to the criterion of resolving
\rg\ for planning future observations.  This approach defines
$v_\mathrm{obs}(r)$ as the line-of-sight rotation velocity of a parcel
of gas at some radius $r$ in a galaxy, and $v_\mathrm{gal}(r)$ as the
rotation speed that parcel would have in the absence of a central BH
(which can be determined from $M_\star(r)$, accounting for uncertainty
in $\Upsilon$).  The figure of merit $\Gamma_\mathrm{FOM}$ is defined
in terms of the confidence level at which $v_\mathrm{obs}$ can be
distinguished from $v_\mathrm{gal} \sin i$ using spatially resolved
observations of disk kinematics, if the uncertainty in measured
velocity at a given location is $\delta v$. The appropriate distance
$r$ to compute this figure of merit is described as the smallest
resolvable distance from the galaxy center, corresponding to the
telescope's beam size, since this is the radius within which the BH
mass is most dominant in the data. 

Our NGC 1332 data highlight a particular issue with this figure of
merit definition. Along the disk major axis at a distance of one
resolution element from the center in NGC 1332, beam smearing and
rotational broadening make the observed line profiles extremely broad
and asymmetric, breaking the one-to-one correspondence between
position and line-of-sight rotation velocity that would be observed in
data of much higher angular resolution.  If $v_\mathrm{obs}(r)$ is
identified with the mean or centroid of the line-of-sight velocity
profile, it will be subject to the beam-smearing effect illustrated in
Figure \ref{beamsmear}, in which case $v_\mathrm{obs}(r)$ can deviate
strongly from the disk's intrinsic $v\sin i$ out to radial scales
significantly larger than the angular resolution of the
observations. The definition of $\Gamma_\mathrm{FOM}$ does not account
for the confounding effects of beam smearing and disk inclination on
the line profiles, an effect that becomes increasingly important at
higher disk inclinations.\footnote{The \citet{davis2014} figure of
  merit is maximized at $i=90\arcdeg$ because this orientation
  maximizes the line-of-sight projection of the disk rotation
  velocity. By this measure, the figure of merit approach predicts
  that BH mass measurements would be most accurate for disks oriented
  exactly edge-on, all else being equal. In fact, the progressive loss
  of information at higher inclination due to beam smearing represents
  a key limiting factor for mass measurement, but the figure of merit
  $\Gamma_\mathrm{FOM}$ does not explicitly incorporate this effect.}
Using the $\Gamma_\mathrm{FOM}$ criterion, \citet{davis2014} argues
that BH masses can be determined even when the observational
resolution is $\sim2$ times larger than the angular size of \rg. We
suggest that observations with such low resolution would not generally
have the power to constrain \mbh\ accurately.

We propose that the best criterion to plan future observations is
simply whether \rg\ is resolved along the disk's projected minor axis
for the anticipated value of \mbh. The ideal situation is one in which
the observations provide several independent resolution elements
across \rg, even along the minor axis of the disk. Such high
resolution observations will have the capability to fully lift
degeneracies between \mbh, $\Upsilon$, and \sigmaturb, and yield
highly accurate (and precise) measurements of \mbh. Deriving a more
rigorous figure of merit criterion for gas-dynamical BH detection
would require a comprehensive suite of simulations incorporating beam
smearing, to test the impact of angular resolution, disk inclination,
S/N, and other parameters on the accuracy of BH mass determination.

\subsection{Model fitting methods and sources of error}

In past gas-dynamical measurements of BH masses using \hst\ data,
models were fit to quantities measured from the data, specifically to
the line-of-sight velocity at each spatial pixel, and sometimes to the
line-of-sight velocity dispersion as well. This is a relatively
time-consuming technique, because it requires measurements of
\vlos\ and \sigmalos\ at each point in the modeled velocity field for
each model realization so that the models and data can be compared. In
principle, the modeled emission-line profiles could be compared
directly with the data instead of going through the intermediate step
of measuring kinematic moments from each model iteration. However,
despite some initial attempts \citep{bertola1998, barth2001conf}, direct
fitting of modeled line profiles has not proved to be a successful
approach for BH mass measurements based on optical emission-line
data. Obstacles to direct line-profile fitting for optical data
include the blending of multiple emission lines such as \hal\ and the
[\ion{N}{2}] $\lambda\lambda6548, 6583$ lines, the presence of broad
emission-line components blended with the resolved narrow-line
emission, and the difficulty of measuring or modeling the
emission-line surface brightness profile to sufficiently high accuracy
for direct line-profile fitting to succeed.

The ALMA data, on the other hand, are much more amenable to direct
fitting of models to the observed data cube. Fitting models to the
data cube makes use of all available information in the data and
should therefore be preferred whenever it is feasible.  We found that
for this dataset, fitting to the PVD (either in whole or over
restricted regions) did not lead to substantial changes in the
inferred \mbh, compared with fitting to the full information in the
data cube. Furthermore, fitting models directly to the data cube is
the most efficient approach, in that it avoids the additional steps of
extracting a PVD or kinematic moment maps from each individual
realization of a modeled data cube for a particular parameter
set.

Gas-dynamical model fits can lead to very high-precision constraints
on \mbh, as discussed by \citet{gould2013}. However, in the regime of
very precise model fitting results, it becomes very important to
explore potential systematic uncertainties that might affect the
measurement, including uncertainty in the slope or shape of the
extended mass profile and uncertainty in the turbulent velocity
dispersion profile. These problems are compounded when the BH's sphere
of influence is not well resolved, or when models are fitted to a
spatial region in which most pixels are at locations well outside of
\rg. Neglecting these systematics could lead to catastrophic errors in
determining \mbh-- that is, errors far larger than the formal
model-fitting uncertainties.  The ideal situation is one in which the
data quality (resolution and S/N) are sufficient to allow model
parameters to be constrained by fitting models only to spatial regions
within $r<\rg$, the region in which the kinematics are maximally
sensitive to \mbh.

In this work, we have focused on modeling uncertainties due to the
degeneracy between rotation and turbulent motion, since this problem
is particularly severe at high disk inclination; in other cases,
uncertainty in the stellar mass profile may be the dominant
contribution to the error budget.  Often, in gas-dynamical model
fitting, the stellar luminosity profile is taken to be a fixed
profile, with just a single scaling factor \ml\ as a free
parameter. This procedure is, however, prone to underestimate the
uncertainty in \mbh, because model fits can converge tightly on a
best-fitting value of \mbh\ even if the measured stellar luminosity
profile deviates from the actual extended mass profile of the
galaxy. When \rg\ is not well resolved, it is particularly important
to incorporate a realistic level of uncertainty in the \emph{slope} or
shape of the deprojected stellar luminosity profile into the model
fitting process. We have not explored this uncertainty in detail in
this work, but our simple experiment of measuring \mbh\ using the
galaxy profile measured from the \hst\ F814W image illustrates the
nature of the problem. For molecular gas-dynamical \mbh\ measurements,
the central light profile of the host galaxy will most often be
severely impacted by dust absorption, and extinction may still be an
issue even in the $K$ band, as illustrated by the VLT SINFONI image of
NGC 1332 from \citet{rusli2011}. This problem can be alleviated to
some extent by masking out the most heavily obscured regions of the
disk when measuring the stellar luminosity profile, or by using
multi-color data to model and remove the effects of extinction.

Additionally, radial gradients in the stellar mass-to-light ratio, if
present, would lead to errors in inferring $M_\star(r)$ from the
galaxy's light profile. The impact of \ml\ gradients on
stellar-dynamical measurements has been investigated by
\citet{mcconnell2013}, who found that plausible \ml\ gradients could
bias BH mass measurements in elliptical galaxies by $\sim20-30\%$. The
influence of \ml\ gradients on gas-dynamical measurements should be
less severe since gas-dynamical data only samples the stellar mass
profile within the region enclosed by the disk's radius, but past
gas-dynamical work has not directly tested or simulated the
contribution of \ml\ gradients to uncertainty on \mbh.  The best and
most secure approach to circumvent these difficulties is to obtain
kinematic data that highly resolve \rg, so that $M_\star(r)$ is much
smaller than \mbh\ over the innermost resolution elements of the data.
Then, uncertainty in the stellar mass profile will not have a large
impact on \mbh.

Another potential source of systematic error in the model fits is the
CO surface brightness map. Beam smearing (both in the data and the
models) mixes information on rotation velocity, turbulent linewidth,
and surface brightness near the disk center, and errors in the assumed
surface brightness map can lead directly to errors in \mbh. The high
inclination of the NGC 1332 disk makes it much more difficult to
discern any surface brightness substructure that may be present. In
general, resolving the emission-line surface brightness structure on
small radial scales is a key requirement for accurate modeling, along
with resolving the kinematic structure. Ideally, the disk's surface
brightness profile should be resolved on scales of $\rg\cos i$ or
smaller.  Given the other large sources of systematic uncertainty in
modeling the NGC 1332 data, we have not explored the impact of surface
brightness errors on our model fits, but unresolved surface brightness
substructure could be responsible for a portion of the high
\chisq\ values we obtain.

As noted by \citet{gould2013}, it is also the case that distance
uncertainty contributes directly to uncertainty in \mbh, because the
value of \mbh\ inferred from dynamical modeling scales linearly with
the assumed distance. Most BH mass measurements have not incorporated
distance uncertainties into their error analysis, but this should be
borne in mind in situations where the formal model-fitting precision
on \mbh\ is so high that distance uncertainty becomes a major
component of the error budget on \mbh.  For NGC 1332, recent distance
measurements listed in the NASA/IPAC Extragalactic Database (NED) are
between 21.9 and 24.6 Mpc \citep{kundu2001, tully2013}, and the
contribution of distance uncertainty to the error budget on \mbh\ is
much smaller than the other sources of systematic uncertainty
associated with the dynamical modeling.

\subsection{Disk structure}

A major issue for gas-dynamical BH mass measurements is the treatment
of the turbulent velocity dispersion. The disk's physical thickness
will depend on $\sigmaturb/\vrot$, where $\vrot$ is the disk's
rotational velocity, and if the turbulent velocity dispersion
contributes an effective dynamical pressure that supports the disk
against gravity, then this must be accounted for in the dynamical
modeling. Enclosed mass scales with $\vrot^2$ for a dynamically cold
disk or $\sigma^2$ for a purely dispersion-supported system, or (very
roughly) $\vrot^2+\sigma^2$ if both rotation and random motions
provide dynamical support for the disk.  Thus, if
$(\sigmaturb/\vrot)^2\ll1$, then the disk can be treated as
dynamically cold. In our model calculations, we have included
\sigmaturb\ merely as an empirical broadening to the emergent CO
linewidths from the disk, but we have not ascribed any dynamical
importance to $\sigmaturb$. These models effectively assume a
dynamically cold, flat disk in which $\vrot = \vcirc$.

\begin{figure}
  \scalebox{0.4}{\includegraphics{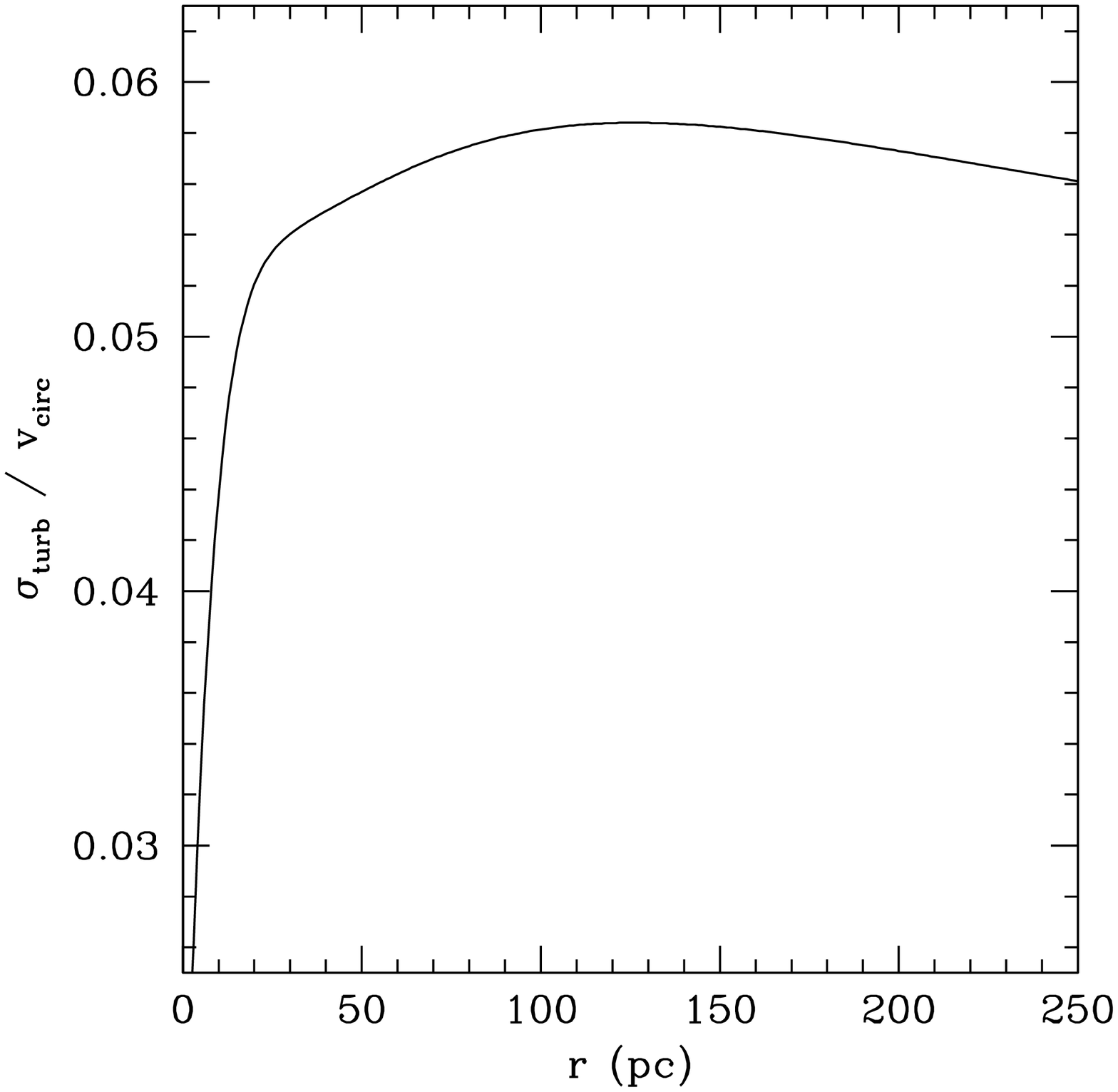}}
  \caption{Radial profile of $\sigmaturb/\vcirc$ for the model with
    $\sigmaturb=24.7$ \kms\ and $\mbh=6.0\times10^8$ \msun.}
  \label{sigmaoverv}
\end{figure}

Figure \ref{sigmaoverv} illustrates the radial variation of
$\sigmaturb/\vcirc$ for the best-fitting model with the flat
\sigmaturb\ profile having $\sigmaturb=24.7$ \kms\ and
$\mbh=6.0\times10^8$ \msun. The maximum value reached by
$\sigmaturb/\vcirc$ is 0.058, at $r=125$ pc. This is a sufficiently
low value to justify the treatment of the disk as thin and dynamically
cold.

In the models with turbulent velocity dispersion gradients, the peak
\sigmaturb\ values are much higher: 268 \kms\ for the exponential
\sigmaturb\ profile and 100 \kms\ for the Gaussian profile. For
reasons described previously, these high \sigmaturb\ values are
probably spurious, since they fail to reproduce the shape of the PVD
at small radii and high velocities.  In these models,
$\sigmaturb/\vrot$ reaches very high values near the disk center,
because with no BH, \vrot\ goes to zero at very small radii. This
would imply that the inner disk becomes very thick and dispersion
dominated, and if this were the case our model assumptions would break
down badly. While this is an extreme and unlikely scenario, we do not
actually have a strong constraint on the maximum value of
\sigmaturb\ in the NGC 1332 disk. A definitive measurement of
$\sigmaturb/\vrot$ across the disk can only be done using data of
higher angular resolution.  For now, we consider the
$\sigmaturb/\vrot$ profile shown in Figure \ref{sigmaoverv} to be a
reasonable estimate, providing some reassurance that the disk can be
modeled as thin and dynamically cold.

Future high-resolution ALMA observations will likely be able to
constrain BH masses tightly by resolving kinematics within \rg.  In
the regime of extremely high-precision BH mass measurement with
well-controlled systematics, the impact of turbulent pressure support
on the derived BH mass might not be negligible in comparison with
other sources of error, in which case other approaches such as Jeans
equation modeling or application of the asymmetric drift formalism
might be necessary.

In this work as in most gas-dynamical BH detections, the disk has been
treated as essentially a surface of zero thickness. In a highly
inclined disk of nonzero thickness, any line of sight through the disk
will pass through regions having different line-of-sight rotation
velocities. If the disk were optically thin but geometrically thick,
it would be important to model the full line-of-sight velocity
profiles through the disk rather than treating the disk as a thin
surface. The NGC 1332 molecular disk is likely to be optically thick
to CO(2--1) emission, as is generally the case in molecular clouds
\citep[e.g.,][]{bolatto2013}, and if it is as thin as suggested by the
$\sigmaturb/\vrot$ profile, modeling it as a thin surface should be a
reasonable approximation to its structure.

The disk also exhibits a mild kinematic twist that signals the likely
presence of a warp, probably similar to the warp in the maser disk of
NGC 4258 \citep{herrnstein2005}. In this first examination of the NGC
1332 disk kinematics, we have not attempted to model the warp, but we
plan to explore warped-disk models in future work. One approach to
modeling a warped disk is to use a tilted-ring model in which the
radial variation of the ring inclination and orientation angles is set
by the disk's measured kinemetry profiles
\citep[e.g.,][]{neumayer2007}.

The high values of \chisq\ found for our best-fitting models, and the
systematic deviations between the modeled line profiles and the data
(Figure \ref{pvdslices}) indicate that there are real and important
aspects of the NGC 1332 disk structure and/or dynamics that are not
incorporated in the models. The largest of these systematic problems
are likely to be the inadequacy of our \sigmaturb\ models and CO
surface brightness model, and the fact that the warp is neglected in
our dynamical models. It is also possible, however, that there may be
real departures from circular rotation in the disk such as $m=2$
perturbations to cloud orbits \citep[e.g.,][]{wong2004}, or localized
departures from circular orbits due to star formation or other
processes within the disk. The line profiles shown in Figure
\ref{pvdslices} show that the largest deviations of the models from
the data are systematic and roughly symmetric on the blueshifted and
redshifted sides of the disk, suggesting that localized random
velocity irregularities are not the dominant contribution to the large
\chisq.

Observations of other transitions from the $^{12}$CO rotational
ladder, as well as $^{13}$CO lines and lines of other molecular species
such as HCN and HCO$^{+}$, can provide much more information on the
temperature and density conditions in ETG disks
\citep[e.g.,][]{crocker2012, bayet2013}, and disks such as the one in
NGC 1332 will be important targets for further ALMA
observations. Additionally, it would be interesting to compare the
kinematics and turbulent velocity dispersion profile of the molecular
gas with the kinematics of ionized gas on the same angular
scales. This can be done with \hst\ STIS observations of the
\hal+[\ion{N}{2}] spectral region or other optical lines, or with
adaptive optics observations in cases where  Br$\gamma$,
[\ion{Fe}{2}], or other near-infrared lines are strong enough to
enable kinematic mapping. Direct comparisons of BH mass measurements
using ionized and molecular gas dynamics in the same galaxy would be
worth pursuing as well, to test whether molecular gas is indeed a more
accurate tracer of circular velocity than ionized gas within the close
environments of supermassive BHs.

\subsection{Comparison with previous \mbh\ measurements}

Since our models do not provide quantitatively satisfactory fits to
the ALMA data cube, we cannot compare our results with the previous
\mbh\ measurements from \citet{humphrey2009} and \citet{rusli2011} in
a rigorous fashion.  However, it is clear that models with \mbh\ in
the range found by \citet{rusli2011}, $\mbh=(1.45\pm0.20)\times10^9$
\msun, lead to far higher \chisq\ values than our best-fitting
models. This is the case for all of the \sigmaturb\ prescriptions that
we have examined, so our provisional \mbh\ estimate appears to be
incompatible with their result. The discrepancy is particularly
intriguing in that we are using the same stellar mass profile measured
by \citet{rusli2011} for their mass modeling. 

The mass model from \citet{rusli2011} implies a total enclosed mass of
$\approx2.4\times10^9$ \msun\ within $r=30$ pc, based on their
best-fitting BH mass and their stellar mass profile with
$\ml_R=7.35$. This gives a circular velocity of 580 \kms\ at $r=30$
pc, much larger than the maximum speed of $\approx480$ \kms\ seen in
the ALMA PVD at $r=30$ pc.  Adopting the $1\sigma$ lower bound to
\mbh\ from \citet{rusli2011}, $1.25\times10^9$ \msun, we would obtain
$\vcirc(30~\mathrm{pc}) = 550$ \kms, still much higher than the
observed outer envelope to the PVD at this radius. In our model fit
with \mbh\ fixed to $1.45\times10^9$ \msun, the mass-to-light ratio
converged to a much lower value of $\ml_R=6.02$ in order to attempt to
fit the disk kinematics at large radii, but still gave a very poor
fit, clearly over-predicting the amplitude of the central
velocity upturn as seen in the lower panel of Figure
\ref{pvdfig-flatsigma}.

The \citet{humphrey2009} measurement of $\mbh =
0.52_{-0.28}^{+0.41}\times10^9$ \msun\ was based on constructing a
mass model for NGC 1332 using the hydrostatic equilibrium of the X-ray
emitting gas as a probe of the gravitational potential. The model
includes components describing the stellar mass distribution of the
galaxy (based on a combination of 2MASS data at large scales and the
\hst\ F814W image at small radii, with the dust disk masked out), the
hot gas profile as measured from \emph{Chandra} observations, the
dark matter halo, and the BH. This method requires the presence of an
X-ray emitting hot interstellar medium in hydrostatic equilibrium, and
to date it has only been applied to a small number of ETGs including
NGC 4261, NGC 4472, and NGC 4649 in addition to NGC 1332
\citep{humphrey2008, humphrey2009}. The X-ray derived BH masses for
NGC 4261 and NGC 4649 are in agreement with masses measured for these
galaxies from ionized gas dynamics \citep[NGC 4261;][]{ferrarese1996}
and stellar dynamics \citep[NGC 4649;][]{shen2010}. For NGC 4472, the
X-ray hydrostatic equilibrium method gives
$0.64_{-0.33}^{+0.61}\times10^9$ \msun\ \citep{humphrey2009} while a
stellar-dynamical analysis finds $\mbh=(2.4-2.8)\times10^9$ \msun\ for
model fits that include a dark matter halo \citep{rusli2013}. A
definitive measurement of \mbh\ in NGC 1332 using higher-resolution
ALMA data can provide a critical test of the X-ray and
stellar-dynamical results.

As a fast-rotating ETG with a high central stellar velocity
dispersion, NGC 1332 bears some similarities to galaxies such as NGC
1277 and NGC 1271 which have been found to contain extremely massive
BHs that are outliers falling well above the $\mbh-\lbul$ correlations
of the general population of ETGs \citep{vandenbosch2012, walsh2015,
  walsh2016, scharwachter2015}. Assuming the BH mass from
\citet{rusli2011} for NGC 1332, \citet{kormendyho} discuss whether the
galaxy is an outlier or not relative to the $\mbh-\mbul$
relationship. The answer hinges on whether NGC 1332 is treated as a
flattened, single-component elliptical galaxy, or a two-component S0
with bulge and disk. If it is a two-component bulge+disk system in
which the elongated portion of the galaxy is considered to be a disk
component, then the BH would be moderately over-massive relative to
its small bulge, but if the galaxy is instead a flattened elliptical
with a dominant bulge component accounting for the majority of the
light, then the BH--bulge mass ratio is within the normal range for
ellipticals. \citet{kormendyho} argue that NGC 1332 is a highly
flattened elliptical, in which essentially all of the light can be
ascribed to the bulge component.  \citet{savorgnan2016} also support
the bulge-dominated interpretation. If the BH mass is close to our
provisional estimate and the value determined by \citet{humphrey2009},
then the galaxy would lie closer to the normal BH--bulge mass ratio
for ETGs, and it would even fall in the low-mass tail of the scatter
distribution in the \msigma\ relation for ETGs \citep[Figure 15
  of][]{kormendyho}.

\subsection{Future prospects for BH mass measurements with ALMA}

ALMA observations have an exciting potential for enabling
gas-dynamical BH mass measurements, but it remains to be seen how
widely applicable this method will be for exploring BH
demographics. For ETGs, the available pool of targets for precision
measurement of \mbh\ will be a small fraction of the overall
population of ETGs. Well-defined circumnuclear dust disks are only
seen in $\sim10\%$ of ETGs. Measuring accurate BH masses in these
galaxies will be most successful when (a) the gas kinematics are
dominated by simple disklike rotation, (b) ALMA observations can
resolve $\rg\cos i$, and (c) the molecular emission-line surface
brightness is high enough that the gas kinematics can be mapped on
scales well within \rg. This last requirement could prove to be a
major limiting factor for ALMA measurements of BH masses; it is not
yet known what fraction of circumnuclear disks will have detectable
emission extending inward to $r<\rg$.  In NGC 1332, the high-velocity
emission provides evidence that the disk extends inward to at least
$r\approx15$ pc. However, if there had been a central ``hole'' in the
molecular disk with a radius much larger than $\sim25$ pc then there
would be no high-velocity upturn in the data and accurate measurement
of the BH mass would not be possible even with higher resolution
observations. Molecular clouds may be easily disrupted in the
immediate environments of massive BHs as a result of the extreme shear
\citep{utomo2015} or due to winds or irradiation from intermittent
accretion-powered nuclear activity. Tidal disruption of molecular
clouds in the close environments of supermassive BHs has also been
suggested as an explanation for the lack of young stellar populations
seen in the innermost few parsecs of typical S0 and early-type spiral
galaxies \citep{sarzi2005}.  As ALMA observes larger numbers of ETGs
having circumnuclear disks, it will be possible to determine the inner
structures of these objects and search for any connections with BH
mass, nuclear activity level, or other properties.

Accurate measurements of host galaxy luminosity profiles in the
near-infrared will be a critically important component of future
gas-dynamical \mbh\ measurements. Ideally, these observations should
have angular resolution at least as high as the ALMA data, and
next-generation extremely large telescopes equipped with adaptive
optics will bring greatly improved capabilities for these
measurements.

Several targets for \mbh\ measurement have already been observed by
ALMA in Cycles 0--2, and additional programs have been approved for
Cycle 3, so the number of galaxies with detections of high-velocity
rotation within \rg\ should begin to increase rapidly in the near
future.  The most efficient way to pursue BH mass measurements with
ALMA will be to continue carrying out initial, quick observations of
galaxies with resolution just sufficient to test for the presence of
rapidly rotating gas within \rg.  Finding evidence of high-velocity
rotation within \rg\ will then justify deeper and higher-resolution
observations that can potentially provide exquisite sampling of the
gas kinematics within \rg. Galaxies for which \mbh\ can be measured to
high accuracy using data that highly resolve \rg\ are extremely
valuable, providing firm anchors to local BH demographics and the
BH-host galaxy correlations, in addition to providing strong evidence
that the central massive objects are indeed likely to be supermassive
BHs \citep{maoz1998}.

In our ALMA programs, we are selecting ETG targets based on the
presence of well-defined circumnuclear dust disks, which give
morphological evidence for dense gas in rotation about the galaxy
center. ALMA will likely be an important tool for gas-dynamical
detection of BHs in spiral galaxies as well, but it is not yet known
what fraction of spiral galaxies contain molecular gas in clean
disk-like rotation on scales of $r<\rg$. The dust-disk selection
method we have used to identify ETG targets for ALMA would not be
applicable to spirals, which typically have more complex, filamentary,
or spiral dust-lane structure \citep[e.g.,][]{martini2003}, but the
growing number of high-resolution ALMA observations of nearby spirals
will make it possible to examine molecular gas kinematics in galaxy
nuclei in far greater detail than has previously been
possible. Recently, \citet{onishi2015} presented dynamical modeling of
ALMA HCN and HCO$^{+}$ kinematics in the SBb galaxy NGC 1097, deriving
a BH mass of $1.40_{-0.32}^{+0.27}\times10^8$ \msun. We note that for
a stellar velocity dispersion of 196 \kms\ \citep{lewis2006}, the
gravitational radius of influence of the NGC 1097 BH would be 15.6 pc
or 0\farcs22, while the Cycle 0 data used by \citet{onishi2015} had a
beamsize of $1\farcs6\times2\farcs2$. This is an order of magnitude
too coarse to resolve \rg, and the observed PVD does not show any hint
of a central  velocity upturn. However, the ALMA PVD does
exhibit velocity structure consistent with regular rotation within the
inner $r<10\arcsec$, an encouraging sign that higher-resolution
observations would have the capability to measure \mbh\ accurately if
the disk kinematics continue to be dominated by regular rotation down
to scales within \rg.

\section{Conclusions}

This paper presents an ALMA CO(2-1) observation of the center of NGC
1332 at 0\farcs3 resolution. We find evidence for a disk in orderly
rotation with evidence for a mild kinematic twist, and a central
upturn in maximum line-of-sight velocity consistent with the expected
signature of rapid rotation around a compact central mass. Although
the quality of the Cycle 2 ALMA data is excellent and the central
upturn in maximum rotation speed provides evidence for a compact
central mass in NGC 1332, the value of the BH mass cannot be tightly
constrained due to severe degeneracies between \mbh\ and other
parameters.  These degeneracies stem primarily from beam smearing and
rotational broadening of the line profiles, because the BH sphere of
influence is slightly unresolved along the disk major axis and very
unresolved along the minor axis.  We find that the BH mass is
degenerate with the turbulent velocity dispersion profile of the disk.
Any chosen prescription for the functional form of the disk's
turbulent velocity dispersion profile can lead to formally tight
constraints on the BH mass, but different choices for how to model the
turbulent velocity dispersion can lead to widely divergent conclusions
for the value of the BH mass.

While we are unable to constrain \mbh\ definitively with the present
ALMA data, models with a flat or exponentially declining
$\sigmaturb(r)$ profile and $\sigmaturb$ limited to values of $\leq50$
\kms\ point to a BH mass in the range $\sim(4-8)\times10^8$ \msun. The
model fits strongly disfavor a BH mass as high as the value determined
by \citet{rusli2011} from stellar-dynamical modeling, $1.45\times10^9$
\msun, but agree well with the value obtained from a hydrostatic
analysis of X-ray emission from the diffuse interstellar medium
\citep{humphrey2009}.

When the BH sphere of influence is not highly resolved, gas-dynamical
model fits can have a tendency towards high-precision formal
constraints on \mbh\ that are not necessarily matched by
correspondingly high accuracy, but which simply reflect the specific
choices and assumptions made in constructing models. This can
potentially result in catastrophic errors in determining \mbh\ (i.e.,
errors that are much larger than the uncertainties in \mbh\ derived
from the usual $\Delta\chisq$ criteria). To obtain
gas-dynamical measurements of \mbh\ that are both precise and
accurate, there is no substitute for observations that \emph{highly}
resolve the BH sphere of influence.

We argue that the appropriate criterion for quantifying the
feasibility of carrying out a gas-dynamical mass measurement is
whether the BH's radius of influence is resolved along the disk's
\emph{minor} axis-- in other words, the angular resolution of the
observation should be smaller than the angular size of $\rg\cos
i$. Observations satisfying this criterion (or better) will resolve
both the essential kinematics and the emission-line surface brightness
substructure in the BH environment, and will lift the degeneracy seen
in this dataset between turbulent and rotational motion in the inner
disk. When $\rg\cos i$ is highly unresolved, dynamical models can
suffer from potentially large systematic uncertainties in deriving
\mbh\ due to uncertainties in the disk's \sigmaturb\ profile and
sub-resolution surface brightness structure, and due to errors in
measurement of the shape of the galaxy's spatially extended mass
distribution. 

This initial observation was designed primarily to test for the
presence of high-velocity emission from within \rg. ALMA is capable of
achieving much higher angular resolution than the 0\farcs3 beamsize of
these data, and new observations of NGC 1332 at 0\farcs04 resolution
have been scheduled for ALMA's Cycle 3. These new observations will
enable much more detailed fitting of dynamical models on scales
$r<\rg$ along the disk's major axis, and will nearly or fully resolve
$\rg\cos i$ (depending on the value of \mbh). We anticipate that the
new data will provide firm constraints on \mbh, demonstrating the full
potential of ALMA for dynamical measurement of BH masses.

\acknowledgements

This paper makes use of data from ALMA program 2013.1.00229.S. ALMA is
a partnership of ESO (representing its member states), NSF (USA) and
NINS (Japan), together with NRC (Canada) and NSC and ASIAA (Taiwan),
in cooperation with the Republic of Chile. The Joint ALMA Observatory
is operated by ESO, AUI/NRAO and NAOJ. The National Radio Astronomy
Observatory is a facility of the National Science Foundation operated
under cooperative agreement by Associated Universities, Inc.  This
research is based in part on observations made with the NASA/ESA
Hubble Space Telescope, obtained from the Data Archive at the Space
Telescope Science Institute, which is operated by the Association of
Universities for Research in Astronomy, Inc., under NASA contract NAS
5-26555. This research has made use of the NASA/IPAC Extragalactic
Database (NED), which is operated by the Jet Propulsion Laboratory,
California Institute of Technology, under contract with the National
Aeronautics and Space Administration.  We thank Timothy Davis and
Martin Bureau for very stimulating discussions on ALMA gas kinematics
during a workshop at Wadham College, Oxford, in March 2015, Jens
Thomas for providing the published stellar mass profile from
\citet{rusli2011}, and the anonymous referee for a helpful report.

LCH acknowledges support from the Chinese Academy of Science through
grant No.\ XDB09030102 (Emergence of Cosmological Structures) from the
Strategic Priority Research Program, and from the National Natural
Science Foundation of China through grant No.\ 11473002.

\emph{Facilities:} \facility{ALMA}, \facility{\emph{HST} (WFPC2)}

\end{document}